\documentstyle[eqsecnum,aps,prd]{revtex}
\begin{document}
\newcommand{\ppp}{\partial}
\newcommand{\V}{\nabla}
\newcommand{\EEE}{\epsilon}
\newcommand{\Lie}{\pounds}
\newcommand{\beq}{\begin{equation}}
\newcommand{\eeq}{\end{equation}}
\newcommand{\beqn}{\begin{eqnarray}}
\newcommand{\eeqn}{\end{eqnarray}}
\newcommand{\nnb}{\nonumber}
\newcommand{\h}[1]{ {\hat{#1}} }
\newcommand{\s}[1]{{\scriptscriptstyle{#1}}}
\newcommand{\ssr}[1]{\scriptscriptstyle{\rm \, #1}}
\newcommand{\stimes}{\mbox{$\bigcirc{\hskip -1.6ex}{\rm s}~\,$}} 
\newcommand{\real}{\mbox{${\rm I\!R}$}}
\newcommand{\removedPt}{\mbox{$\scriptscriptstyle{RP}$}}
\newcommand{\bri}{\mbox{$\breve\imath^{\ssr +}$}}
\newcommand{\1}{{}^{^{(1)}}{\hskip-2.8pt}}
\newcommand{\2}{{}^{^{(2)}}{\hskip-2.8pt}}
\newcommand{\upp}[1]{{}^{\scriptscriptstyle{(#1)}}{\hskip-2.4pt}}
\newcommand{\up}[1]{{}^{^{#1}}{\hskip-3.0pt}}
\newcommand{\base}[2]{ {\scriptstyle (} #1_{\! \ssr{#2}} {\scriptstyle )} }
\newcommand{\bm}[1]{ \mbox{\boldmath{$#1$}} }    
\newcommand{\tfrac}[2]{ \textstyle\frac{#1}{#2} }
\newcommand{\lmk}{\left(}
\newcommand{\rmk}{\right)}
\newcommand{\lkk}{\left[}
\newcommand{\rkk}{\right]}
\newcommand{\lnk}{\left\{} 
\newcommand{\rnk}{\right\}}
\thispagestyle{empty}
\thispagestyle{empty}
{\baselineskip0pt
\leftline{\large\baselineskip16pt\sl\vbox to0pt{\hbox{Department of Physics} 
               \hbox{The University of Tokyo}\vss}}
\rightline{\large\baselineskip16pt\rm\vbox to20pt{\hbox{UTAP-270}
                \hbox{RESCEU-33/97}
                \hbox{gr-qc/xxxxx}           
               \hbox{\today}
\vss}}%
}
\vskip15mm
\begin{center}
{\large\bf Timelike Infinity and Asymptotic Symmetry}
\end{center}

\begin{center}
{\large Uchida Gen} \\
\sl{Department of Physics, The University of Tokyo, Tokyo 113, Japan}
\end{center}
\begin{center}
{\large Tetsuya Shiromizu} \\
\sl{Department of Physics, The University of Tokyo, Tokyo 113, Japan \\
and \\
Research Center for the Early Universe(RESCEU), \\ The University of Tokyo, 
Tokyo 113, Japan
}
\end{center}

\begin{abstract} 

By extending Ashtekar and Romano's definition of spacelike
infinity to the timelike direction, a new definition of asymptotic
flatness at timelike infinity for an isolated system with a source 
is proposed. 
The treatment provides unit spacelike 3-hyperboloid 
timelike infinity and avoids the introduction of the troublesome differentiability
conditions
which were necessary in the previous works on asymptotically flat spacetimes
at timelike infinity. Asymptotic flatness is characterized
by the fall-off rate of the energy-momentum tensor at timelike
infinity, which makes it easier to understand physically 
what spacetimes are investigated.
The notion of the order of the asymptotic flatness 
is naturally introduced from the rate.
The definition gives a systematized picture of hierarchy in the
asymptotic structure, which was not clear in the previous works.
It is found that 
if the energy-momentum tensor falls off at a rate faster than $\sim t^{-2}$,
the spacetime is asymptotically flat and asymptotically stationary 
in the sense that the Lie derivative of the metric with respect to $\ppp_t$
falls off at the rate $\sim t^{-2}$. It also admits
an asymptotic symmetry group similar to the Poincar\'e group.
If the energy-momentum tensor falls off at a rate faster than $\sim t^{-3}$,
the four-momentum of a spacetime may be defined. 
On the other hand, angular momentum is defined 
only for spacetimes in which the energy-momentum tensor falls off at
a rate faster than $\sim t^{-4}$.
\end{abstract}
\vskip1cm




\section{Introduction}
If gravitational collapse takes place in an asymptotically flat
spacetime, we naively expect that the energy of fields and particles 
in the exterior region would be either radiated away to infinity 
or fall into the black hole and that 
the spacetime would approach a vacuum spacetime.
Because the lost energy cannot be restored,
the dynamical processes would eventually die away and the spacetime
would settle down to a stationary state.
Thus, we expect that the final state of gravitational collapse 
be described by a Kerr black hole, 
since the uniqueness theorem states that 
the Kerr spacetime is the unique 
asymptotically flat, stationary, vacuum solution~{\cite{MH}}.
This is the reason why the Kerr solution is used to analyze the black hole
candidates observed in the universe.
In this paper, to confirm the naive picture above,
we shall closely examine the asymptotic structure
of the very late time regime of a spacetime, i.e., timelike infinity,
and see that an ``asymptotic'' vacuum state actually induces 
an ``asymptotic'' stationary state. This approach may be more reasonable 
than the assumption of the exact stationarity
in the sense that {\em exact} vacuum or {\em exact} stationary
state would not be achieved in a real gravitational collapse.

A solution of the Einstein equation is normally given in terms of the
metric components with respect to some specific coordinates. 
However, it is not easy from the component expression
to grasp what physical system the solution actually represents.
One way to overcome the difficulty is to introduce a certain physical index
that characterizes the system. For example, if the energy of the system at a given time
is defined, we can picture the evolution of the system to some extent
from the change in its value.
Moreover, if the multipole moments of a spacetime are defined, 
we may obtain a precise Newtonian analog
by constructing a Newtonian potential having the same moments.
The analog could serve as a powerful guide for the physical
interpretation of the solution. 
Because the gravitational field itself
acts as its own source, we expect
multipole moments can be defined only in spacetimes with 
gravitational fields satisfying
a suitable fall-off condition.
It is shown that {\em stationary} spacetimes are 
sufficiently asymptotically flat for the multipole moments 
to be well defined~\cite{RH}. 
However, there are no arguments on more general spacetimes.
In this paper, as a first step towards defining multipole moments 
of general asymptotically flat spacetimes, 
we explore timelike infinity of a spacetime to seek what fall-off condition
the gravitational field must satisfy so that 
four-momentum or angular-momentum may be defined.

There are some previous works on timelike infinity.   
A definition of timelike infinity 
with {\em everywhere-smooth} unphysical metric
is proposed by Cutler in~{\cite{CC}}. 
The high differentiability condition results 
in demanding that the spacetime have 
vanishing final Bondi mass at timelike infinity. 
This means the definition applies only for spacetimes with no
bounded sources such as black holes and stars.
On the other hand, Porill's definition of the asymptotic flatness at 
timelike infinity is applicable to spacetimes with bounded sources~{\cite{JP}}.
However, as a result of the ``tieing'' 
of the late time regime with the null regime,
the treatment compresses the late time regime, that has a finite volume, 
down to a point.
As a result, we cannot demand smoothness or even differentiability
of the unphysical metric at the point. 
Instead, we need to formulate a notion of 
direction-dependent ``differentiability''
and demand that the metric possess this awkward property.
The treatment makes it  
difficult to understand what kinds of spacetimes are investigated.

Recently, Ashtekar and Romano have formulated an elegant treatment of 
spacelike infinity in a coordinate-independent manner\cite{AR}. 
In their work, the corresponding ``awkward'' differentiability at spacelike
infinity is avoided by regarding the late time regime of a spacetime
not as a point but as a 3-manifold.
This is achieved by abandoning the idea of 
``tieing'' the late time regime and the null regime together.
As a result, there is no necessity to use a conformally completed
unphysical metric for the description of 
the gravitational field. Instead, 
a normal vector field and an intrinsic metric on the timelike slices
foliating the unphysical spacetime are defined, and 
they provide the 3-manifold timelike infinity.

In the present paper, 
we modify the Ashtekar and Romano's treatment of spacelike infinity
to the timelike direction  and investigate
the asymptotic structure of spacetimes around timelike infinity. 
Our investigation is more or less a direct application 
of their work but differs in a number of important points,
which enable us to discuss 1) the asymptotic vacuum state and the
asymptotic stationary state of a spacetime
to explore the final product of gravitational collapse,
and 2) the order of asymptotic flatness to explore how flat the
spacetime must be in order to define physical indices.
First, we take a different view on 
how the fall-off condition on the energy-momentum tensor should be imposed.
We impose conditions on the {\em physical} components of the physical 
energy-momentum tensor, while they impose conditions on the {\em unphysical} 
components. 
This change gives a picture of hierarchy 
in the asymptotic structure at timelike infinity and helps us 
to understand how the structure is systematized, 
which is a point that was obscure in the previous treatments.
Second, a notion of asymptotic Killing fields is introduced.
The formulation of the order of asymptotic symmetries is possible
only by the introduction of such a notion. 
This is also crucial in examining and specifying 
how fast a spacetime becomes stationary.
Also, special attention is paid to a bounded source that 
persists even in the very late time regime of a spacetime, 
in contrast with the spacelike direction. 

The rest of the paper is organized as follows. In
sec.~\ref{sec:Minkowski}, we construct timelike infinity 
with certain desirable features for Minkowski spacetime. 
Following the completion of Minkowski spacetime,
we propose a definition of
asymptotic flatness at future timelike infinity for 
a general spacetime with a source in sec.~\ref{sec:definitions}.
In sec.~\ref{sec:zero}, we examine the fundamental asymptotic structure of spacetimes 
that possess the defined asymptotic flatness.
In sec.~\ref{sec:symmetry}, 
we propose an asymptotic symmetry group which preserves the studied structure. 
A notion of asymptotic
Killing fields is introduced in 
sec.~\ref{sec:translation}, and its relation to the fall-off of
the energy-momentum tensor is clarified.
In sec.~\ref{sec:fields}, we examine the asymptotic behavior of the gravitational fields.
In sec.~\ref{sec:conserved},
we define conserved quantities associated with the asymptotic symmetry. 
The appendices give some examples of asymptotically flat spacetimes with
timelike infinity defined in this paper and the flat Friedmann spacetime 
which is not asymptotically flat as a counter example. 
Throughout the paper, we follow 
the notation of Wald\cite{RW}.


\section{Completion of the Minkowski Spacetime}\label{sec:Minkowski}

In this section, we complete 
Minkowski spacetime~($\hat{\cal M}$,$\hat{\eta}_{ab}$) 
in such a way that future timelike infinity is represented 
not as a point but as a 3-manifold. 
The construction motivates the general definition of asymptotic flatness in the next section.
Hereafter, we only consider {\em future} timelike infinity and thus omit the word future.
All the discussion presented here is also applicable 
to past timelike infinity with minor changes.

In static spherical coordinates~$(t,r,\theta,\phi)$,
the metric of Minkowski spacetime takes the form
%
\beq
 \hat{\eta}_{ab} =  - (dt)_a(dt)_b + (dr)_a(dr)_b +  r^2(d{\sigma})_{ab}
\eeq
%
where 
$(d{\sigma})_{ab}:=
(d\theta)_a(d\theta)_b+\sin^2\theta(d\phi)_a(d\phi)_b$
is the unit 2-sphere metric.
Introduce the standard hyperbolic coordinates~($\rho$,$\chi$,$\theta$,$\phi$)
in the {\em interior} of the future light cone at the origin: 
$t = \rho \cosh \chi $ and 
$r = \rho \sinh \chi$, where $\chi\in [0,\infty)$ and $\rho\in (0,\infty)$. 
In this chart, timelike infinity 
is specified by 
the limit surface $\rho\to\infty$ and the metric takes the form 
%
\beq
      \hat{\eta}_{ab} =  - (d\rho)_a(d\rho)_b + \rho^2 h_{ab}
\eeq
%
where $h_{ab}:=(d\chi)_a(d\chi)_b+{\sinh}^2\chi(d\sigma)_{ab}$ 
is the unit spacelike 3-hyperboloid metric.
If a new coordinate $\Omega:=\rho^{-1}$ is introduced and 
$\h{\cal M}$ is extended to include the points on which $\Omega=0$,
timelike infinity is specified by the surface $\Omega=0$ and 
the metric is found to be singular there: 
%
\beq
     \hat{\eta}_{ab}  =  - \Omega^{-4}(d\Omega)_a(d\Omega)_b 
                 + \Omega^{-2} h_{ab}.             
\eeq
%
To complete the spacetime conformally,
we take a conformal factor of $\Omega^4$ and obtain a
non-singular unphysical metric:
%
\beq
     \eta_{ab} := \Omega^4\hat{\eta}_{ab} =   - (d\Omega)_a(d\Omega)_b 
                             + \Omega^2 h_{ab}.   
\eeq
%
With respect to the metric, 
timelike infinity $\Omega =0$ is a point with no volume.
This is because the components of the metric normal to 
and those tangent to 
the $\Omega$-const$.$ surfaces  are 
rescaled by the same power of $\Omega$ although they diverge with different
powers. In other words, if we 
rescale them according to the power of their divergence,
we obtain timelike infinity that is not compressed to a point. Thus, define
%
\beq
     n^{a}  :=  \Omega^{-4}{\hat \eta}^{ab}(d\Omega)_b            
              =   -(\partial_{\Omega})^a               
\eeq
%
to the direction normal to the $\Omega$-const$.$ surfaces and
%
\beq
     q_{ab}  :=  \Omega^2 \Bigl(\hat{\eta}_{ab} 
                  + \Omega^{-4}F^{-1}(d\Omega)_a(d\Omega)_b\Bigr) 
             =  h_{ab},                          
     \quad \mbox{where}\quad F:=-\Lie_n \Omega=1
\eeq
%
to the direction tangent to the $\Omega$-const$.$ surfaces.
The information on the gravitational fields is completely given by the pair
$(n^a,q_{ab})$, and we do not need the conformally rescaled unphysical metric 
$\eta_{ab}$. With respect to this pair, timelike infinity is 
a 3-surface whose intrinsic metric is $h_{ab}$
and whose normal is $\ppp_{\Omega}$. 
That is, timelike infinity is a unit spacelike 3-hyperboloid.

Note here that the completion presented here does not conserve the causal structure of 
the physical spacetime, in contrast with the conformal completion. 
Hence, it is inappropriate, for example, to discuss the relation of timelike infinity
with null infinity using the method. However, the method provides 
an adequate basis to investigate asymptotic structure of a spacetime 
at timelike infinity.


\section{Definitions}\label{sec:definitions}
In this section, we introduce a definition of asymptotic flatness 
at timelike infinity to order $n$ and define some tensor fields that are useful
for the coming discussion.

We have demonstrated the completion of Minkowski spacetime in the previous section.
It is important to note that the way which the completion is accomplished 
crucially depends on the structure of Minkowski spacetime at timelike infinity. 
This motivates us to define an asymptotic flat spacetime 
as a spacetime which can be completed ``asymptotically'' in a similar
fashion to Minkowski spacetime.
More precisely, a spacetime is said to be asymptotically flat 
if the spacetime can be extended to enclose the points of
infinity and if it is possible to construct the pair $(n^a,q_{ab})$
that has a smooth limit to the infinity.

\medskip

\noindent
{\bf DEFINITION:} \enskip A physical spacetime ($\h{\cal M}$, $\h g_{ab}$) is
said to possess an {\em asymptote at future timelike infinity $\bri$ to order n}
({\sc afti}-$n$) for a non-negative integer $n$,
if there exists a manifold $\cal M$ with boundary $\cal H$, 
a smooth function $\Omega$ defined on $\cal M$, 
and an imbedding $\Psi$ of an open subset $\h{\cal F}$ in $\h{\cal M}$
to $\cal M - \cal H$
satisfying the following conditions:

(1) $\bri:=\ppp{\cal F}\cap({\cal M}-\Psi(\h{\cal M}))$ is not
empty and $\bri\in I^+({\cal F})$ where ${\cal F}:=\Psi({\h{\cal F}})$;

(2) $\Omega\breve{=}0$ and $\V_{a}\Omega\breve{\neq}0$, 
where $\breve{=}$ denotes the equality evaluated on $\bri$;

(3) $n^a:=\Omega^{-4}\Psi^*\h g^{ab}\V_b\Omega$ and 
    $q_{ab}:=\Omega^2(\Psi^*\h g_{ab}+\Omega^{-4}F^{-1}\V_a\Omega\V_b\Omega)$
admit smooth limits to $\bri$ with $q_{ab}$ having signature(+++) on $\bri$,
where $F:=-\Lie_n\Omega$; and

(4) $\lim_{\rightarrow\bri}\Omega^{-(2+n)}\h T_{\h\mu\h\nu}=0$
where $\h T_{\h\mu\h\nu}:=\Psi^*[\base{\h e}{\mu}^a\base{\h e}{\nu}^b\h T_{ab}]$ 
in which $\h T_{ab}$ is the physical energy-momentum tensor and
$\{ \base{\h e}{\mu}^a \}_{\ssr{\mu=0,1,2,3}}$ 
is a tetrad of $(\h{\cal M},\h g_{ab})$.
 
\medskip

Let us now discuss the meaning of the conditions.
First of all, note that the asymptotic flatness is defined for an 
open {\em subset} $\h{\cal F}$ of the spacetime manifold.
This is because isolated bodies may be present at the late time regime of a 
spacetime and thus the {\em whole} spacetime is not expected to be 
asymptotically flat. 
If there are isolated bodies at the late time regime, 
the region excluding the bodies should be chosen as $\h{\cal F}$.
The former part of the first condition ensures that 
the infinity $\bri$ consists of points that are included 
by extending the physical manifold. 
Such $\bri$ resides not in the region $\Psi(\hat{\cal M})$, 
where the physical manifold is mapped, but on the boundary 
of the region of interest, $\ppp{\cal F}$.
The latter part assures that $\bri$ exists
in the {\em future} of the physical spacetime
so that $\bri$ represents {\em future} timelike infinity.
The second condition assures that $\bri$ is infinitely far away.
The third demands that the asymptotic spacetime can be completed
asymptotically in a similar fashion to the case of Minkowski spacetime 
and that the infinity $\bri$ is {\em timelike}.
This means that $\Omega$ asymptotically plays a role of $t^{-1}$
in Minkowski spacetime.  
The fourth condition specifies how the physical components
of the physical energy-momentum tensor should decay.
This condition is different from the corresponding one in the
definition given by Ashtekar and Romano~\cite{AR}.
If we literally translated their prescription for spacelike infinity
to our timelike infinity, we would demand that the energy-momentum tensor
fall off faster than $\Omega^1$, $\lim_{\to\bri}\Omega^{-1}\Psi^*\h T_{ab} = 0$,
and consider such terms as ${q_a}^m {q_b}^n\Omega^{-1}\Psi^*\hat{R}_{mn}$
vanish on $\bri$.
Although it is not explicitly stated in\cite{AR},
this means that the energy-momentum tensor is evaluated by 
the components with respect to a {\em unphysical} coordinate basis,
$\base{e}{\mu}^a\base{e}{\nu}^b\Psi^*\h T_{ab}$
where $\{ \base{e}{\mu}^a \}_{\ssr{\mu=0,1,2,3}}$
is a tetrad of the unphysical spacetime. 
Here, we take an alternative view. 
We evaluate the {\em physical} energy-momentum tensor 
by its components with respect to a {\em physical}
coordinate basis, 
$\h T_{\h\mu\h\nu}=\Psi^*[\base{\h e}{\mu}^a\base{\h e}{\nu}^b\h T_{ab}]$.
In this view, it is found that 
the richness of asymptotic structures that a spacetime possesses
depends on how fast the energy-momentum tensor falls off in the
spacetime. Thus, we need to specify the rate of the fall-off
when defining asymptotic flatness.
To evaluate $\h T_{\h\mu\h\nu}$ hereafter, 
we use a tetrad consisting of a unit normal vector 
to $\Omega$-const$.$ surfaces, ${\hat n}^a$,
and the triad on $\Omega$-const$.$ surfaces, 
$\{ \base{\hat e}{I}^a \}_{{\rm I}=1,2,3}$. They satisfy
%
\beqn
        \hat{g}^{ab} &=& - \hat{n}^{a}\hat{n}^{b}
                          + \base{\hat e}{I}^a \base{\hat e}{I}^b 
                                                     \label{eq:g=nnee}\\
   \Psi^*\hat{n}^a\, &=& \frac{n^a} { \sqrt{ -\Psi^*\hat{g}_{ab} n^a n^b } } 
                      =  \Omega^{2}F^{-1/2}n^{a},    \\
  \Psi^*\base{\hat e}{I}^a &=& \Omega \,\base{e}{I}^a      \label{eq:^e=Oe}
\eeqn 
%
where $\{ \base{e}{I}^a \}_{{\rm I}=1,2,3}$ is the smooth triad of
$q_{ab}$. Eq. (\ref{eq:^e=Oe}) will be obtained below.
Since all the equations appearing in the following discussion
are those on $\cal M$, 
unless it may cause ambiguity,
we omit hereafter
$\Psi^*$ in front of the tensors defined on 
$\h{\cal M}$ for brevity.

It is meaningful to note here that a scalar field $\phi$
in the Schwarzschild spacetime
decays asymptotically at the rate $\sim t^{-2}$ or faster, according to the
Price's theorem\cite{PT}. Thus, the Schwarzschild spacetime 
with such a scalar field is asymptotically flat to order 1. 

Now consider the Schwarzschild spacetime.
We can take the region $r>2m$ for ${\cal F}$. Then, all the conditions
above are satisfied and $\bri$ is represented by 
a unit spacelike 3-hyperboloid with a point removed from it.
(See appendix A for the details.)
The removed point corresponds to the future of the region $r\leq 2m$, or the
region excluded from $\h{\cal M}$. Recalling that  a unit spacelike
3-hyperboloid has a topology of $S^2\times\real^+$, we see that the
topology of $\bri$ is now $S^2\times(\real^+ -\{p\})$ 
where $\real^+$ denotes $[0,\infty)$.
This result may be regarded as a common feature of the spacetimes with sources
and motivates the following definition.

\medskip

\noindent
{\bf DEFINITION:} A spacetime ($\hat{\cal M}$, $\hat{g}_{ab}$) is
said to be {\em asymptotically flat at future timelike infinity with a 
source to order n} ({\sc afftis}-$n$) 
if ($\hat{\cal M}$, $\hat{g}_{ab}$) possesses an asymptote at timelike infinity to order $n$
and $\bri$ has topology of $S^2\times(\real^+ -\{p\})$.

\medskip

For the convenience of the following discussion, 
we define some useful tensors and show the relations among them. 
From the definition, 
the physical metric $\hat{g}_{ab}$ and its inverse ${\hat g}^{ab}$ 
are given by
%
\beqn
     \hat{g}_{ab} &=& -\Omega^{-4}F^{-1}\nabla_{a}\Omega\nabla_{b}\Omega
                          +\Omega^{-2}q_{ab}  \\
     \hat{g}^{ab} &=&   - \Omega^{4}F^{-1}n^{a}n^{b}+\Omega^{2}q^{ab}
               \label{eq:gINV}
\eeqn
%
respectively with functions and tensors that are smooth in $\cal F$ and on $\bri$.
From eqs.(\ref{eq:gINV}) and (\ref{eq:g=nnee}),
we obtain eq.(\ref{eq:^e=Oe}).
A smooth projection operator on the $\Omega$-const$.$ surfaces 
can be defined from smooth $q^{ab}$ and $q_{ab}$:
%
\beqn
     {q^{a}}_{b} & := & q^{ac}q_{cb} \nonumber \\
                 & = & {\delta^{a}}_{b} + F^{-1}n^{a}\nabla_{b}\Omega.
\eeqn
%
The operator is the same as the one defined from $\h q^{ab}$
and $\h q_{ab}$. Thus, derivative operators $D_a$ and $\h D_a$ 
that are associated with $q_{ab}$ and $\h{q}_{ab}$ respectively do not differ.
Hence,
%
\beq
   {\up 3}\hat{R}_{abc}{}^d = {\up 3}{R_{abc}}^d \quad \mbox{and} \quad
    {\up 3}\hat{R}_{ab}={\up 3} R_{ab}
\eeq
%
where $(\hat{D}_a\hat{D}_b-\hat{D}_b\hat{D}_a)\,\omega_c =: 
                    {\up 3}\hat{R}_{abc}{}^d\omega_d$,  
${\up 3}\hat{R}_{ab}:={\up 3}\hat{R}_{acb}{}^c$,
$(D_a D_b - D_b D_a)\,\omega_c =: {\up 3}{R_{abc}}^d\omega_d$ 
and ${\up 3}R_{ab}:={\up 3}{R_{acb}}^c$. On the other hand, the physical extrinsic curvature 
of the $\Omega$-const$.$ surfaces does not admit a smooth limit to $\bri$:
%
\beqn
      \hat{K}_{ab} & :=  & \frac12\Lie_{\h n}\h q_{ab} 
                                                           \nonumber \\
                   &  =  & \Omega^{-1}F^{1/2}q_{ab} + \frac{1}{2}F^{-1/2}
                             \Lie_{n}q_{ab}.
\eeqn 
%
Thus, define a smooth tensor
%
\beqn
      K_{ab} & :=  & \Omega\hat{K}_{ab} \nonumber \\
             &  =  & F^{1/2}q_{ab} 
                     + \frac{1}{2}\Omega F^{-1/2}\Lie_{n}q_{ab}.
\eeqn 
%
Hereafter, we raise and lower 
the indices of tensors without hats on $\Omega$-const$.$ surfaces
by the smooth metric $q_{ab}$ and the inverse $q^{ab}$.  For example,
${K^a}_b:=q^{ac}K_{cb}$.


\section{ The zero-th order Asymptotic Structure}\label{sec:zero}
In this section, we examine the asymptotic structure
of an {\sc afti-0} spacetime using the Einstein equation.

Consider the momentum constraint equation expressed in terms of smooth
tensors and $\h T_{\h\mu\h\nu}$:
%
\beqn
     \Omega^{-2} \base{\h e}{I}^a \h n^{b}\h T_{ab} 
        = [ D_{c}K^{c}_{a}-D_{a}K ] \base{e}{I}^a.
\eeqn
%
In {\sc afti-0} spacetimes, the left-hand side vanishes on $\bri$
because of the fall-off condition on the energy-momentum tensor.
Therefore, we get
%
\beq
     D_c K_a{}^c - D_a K \breve= 0.            
\eeq
%
Substituting eq.(3.9) into this equation, we find
%
\beq
     F \breve=\mbox{const}. \label{eq:F=const.}
\eeq
%
By virtue of the conformal freedom of the function $F$,
the above constant can be set to unity without any loss of generality.
To see this, complete the spacetime with a function $\Omega':=\alpha\Omega$
instead of $\Omega$, where $\alpha$ is a smooth function
that is constant on $\bri$. 
Then, the function $F':=-\Lie_{n'}\Omega'$ in this completion
is given by $F'\breve=\alpha^{-2}F$.
Thus, there is always a function $\Omega'$ that induces $F'\breve=1$
for any given function $\Omega$.
We call a completion with such a function a {\em unit hyperboloid completion}
for the reason that will be clarified below. Because there is no loss of
generality, hereafter we consider only unit hyperboloid completions.
In the completion, 
%
\beq
     F = 1 + \sum_{\ell=1}^\infty \up{(\ell)}F \cdot \Omega^\ell,
                                \label{eq:unitF}
\eeq
%
where $\up{(n)}F$ are defined by
%
\beq
     \1 F := \lim_{\to\bri} \Omega^{-1}(F-1)
       \hspace{3ex} \mbox{and} \hspace{3ex}
     \up{(n)} F := \lim_{\to\bri} \Omega^{-n}
                        (F-1-\sum_{\ell=1}^{n-1}\up{(\ell)} F)
       \hspace{3ex} \mbox{for $n\geq 2$};
\eeq
%
and 
%
\beq
     K_{ab} = q_{ab} + \frac{1}{2}\Omega (\, \1 F q_{ab} 
            + \Lie_{n}q_{ab} ) + O(\Omega^2). \label{eq:unitKab}
\eeq
%

Now consider the space-space components of the Ricci tensor: 
%
\beqn
     \Omega^{-2} \base{\hat e}{I}^a \base{\hat e}{J}^b \hat{R}_{ab} 
           = [\, {\up 3}R_{ab} & + &  KK_{ab} - F^{1/2}K_{ab} 
               - 2K_{ac}K^{c}_{b}  \nonumber\\
                & & + \Omega F^{-1/2}\Lie_{n}K_{ab}
               +F^{-1/2}D_{a}D_{b}F^{1/2} ] \base{e}{I}^a \base{e}{J}^b.
             \label{eq:Ree}
\eeqn
%
In {\sc afti-0} spacetimes, the left-hand side vanishes on $\bri$
by virtue of the fall-off condition on the energy-momentum tensor. 
Substituting eqs. (\ref{eq:unitF}) and (\ref{eq:unitKab}) into eq.(\ref{eq:Ree}),
we obtain
%
\beq
     {\up 3}R_{ab} \breve= -2 q_{ab}. \label{eq:original}
\eeq
%
Here, recall that $\bri$ is a sub-manifold of $\cal M$ and thus
there is an identity imbedding $\Pi$ of $\bri$ in $\cal M$.
Using the map $\Pi$, a tensor field $A_{ab}$ defined on $\cal M$ 
induces a tensor field $\bm A_{ab}:=\Pi^* A_{ab}$ on $\bri$.
(Hereafter, we denote the tensor fields on the manifold $\bri$ in boldface.)
It is important to note here that 
if a tensor field is tangential to $\bri$, 
or $q_a{}^m q_b{}^n A_{mn}\breve=A_{ab}$,
the induced tensor field contains the same information as
the original. Hence, the investigation of an equation on $\bri$ with 
tensor fields tangential to $\bri$ can be done by considering the
induced equation on $\bri$.
Thus, let us consider the following equation induced by  eq.(\ref{eq:original})
on $\bri$.
%
\beq
     {\up 3}\bm R_{ab} \breve= 
                    - 2\bm q _{ab}.
\eeq
%
The equation tells us that
the unphysical 3-spacetime $(\bri,\bm q_{ab})$ has negative constant curvature:
%
\beq
     {\up 3}\bm{R}_{abcd} 
       \breve= -2\bm q_{a[c}\bm q_{d]b}. \label{eq:constCRV}
\eeq
%
In other words, timelike infinity $(\bri,\bm q_{ab})$ is isometric 
to a unit spacelike hyperboloid in Minkowski spacetime and 
$\bm q_{ab}\breve=\bm h_{ab}$. 
Note here that it is isometric to a {\em unit} hyperboloid
because the {\em unit} hyperboloid completion induces $F\breve=1$.
This is where the name comes from.
 
The fall-off condition on the time-time component of the
energy-momentum tensor gives nothing new. 

Now that we have explored everything that can be derived
from the fall-off condition which {\sc afti-0} spacetimes satisfy,
the asymptotic structure of {\sc afti-0} spacetime can be summarized as 
follows.

\medskip

\noindent
{\bf DEFINITION:} 

The function $F$ which is unity on $\bri$, i.e., $F\breve=1$ and 
the manifold $\bri$ with normalized negative-constant curvature,
i.e., $\bm q_{ab}\breve=\bm h_{ab}$ are called 
the {\em zero-th order asymptotic structure} of
an {\sc afti}-$0$ spacetime.

\medskip

The zero-th order asymptotic structure corresponds to 
the universal structure of {\sc am} spacetimes in {\cite{AR}}. 
However, they differ in that the former does not contain 
the equivalence class of $\Omega$ 
constructed by regarding two completions $\Omega$ and $\Omega'$ 
equivalent if and only if they satisfy
\beq
  \lim_{\to\bri} \frac{\Omega' - \Omega}{\Omega^2} = 0
      \label{eq:eqvclass}
\eeq
while the latter does.
The reason is that the equivalence class does not arise on its own
from the definition of an {\sc afti-0} spacetime. This observation is also
supported by the discussion of asymptotic symmetries in
section \ref{sec:symmetry}.

Recalling that 
the difference between an {\sc afti-0} spacetime and an {\sc afftis-0} spacetime
is the topology of $\bri$, we define the asymptotic structure of 
an {\sc afftis-0} spacetime as follows.

\medskip

\noindent
{\bf DEFINITION:} 

The function $F$ which  is unity on $\bri$, i.e., $F\breve=1$ and 
the manifold $\bri$ with normalized negative-constant curvature,
i.e., $\bm q_{ab}\breve=\bm h_{ab}$ and topology of $S^2\times(\real^+ -\{p\})$
are called  the {\em zero-th order asymptotic structure} of
an {\sc afftis}-$0$ spacetime.

\medskip

Let us consider the form of the physical metric $\h g_{ab}$ 
in an {\sc afti-0} spacetime to understand 
the zero-th order asymptotic structure more.
From the fact that an {\sc afti-0} spacetime possesses
the zero-th order asymptotic structure,
we see that the expansions of the function $F$ and the tensor field
$q_{ab}$ are given by
%
\beqn
       F &=& 1+O(\Omega) \nnb\\
  q_{ab} &=& h_{ab} +f(d\Omega)_a(d\Omega)_b 
             + f^{\ssr{I}}(d\Omega)_{\s(a}\base{e}{I}_{b\s)}
             +O(\Omega) \nnb
\eeqn
%
in an {\sc afti-0} spacetime
where $f$ and $f^{\ssr{I}}$ are smooth functions.
With the expansion above, 
the definitions of $F$ and $q_{ab}$ give
%
\beq
   \h g_{ab}= - \Omega^{-4}[ 1+O(\Omega) ](d\Omega)_a(d\Omega)_b
              + \Omega^{-2} \left[
                       h_{ab}  +f(d\Omega)_a(d\Omega)_b 
                   +f^{\ssr{I}}(d\Omega)_{\s(a}\base{e}{I}_{b\s)} +O(\Omega)
                           \right].\label{eq:Xpanded}
\eeq
%
Introducing a conformal time $\eta:=\ln\Omega$,
we obtain
%
\beq
    \h g_{ab}= \upp0\h g_{ab}+ \Omega\upp1\h g_{ab}+\cdots \label{eq:gab=...}
\eeq
%
where
%
\beq
   \upp0\h g_{ab}:=  (e^{-\eta})^2 \left[
                                        - (d\eta)_a(d\eta)_b + h_{ab}
                                  \right]\label{eq:(0)gab=}
\eeq
%
and $e^{2\eta}\cdot\upp0\h g_{ab}, e^{2\eta}\cdot\upp1\h g_{ab},\cdots$
are tensor fields that do not depend on $\Omega$.
If we transform the coordinates, it can be seen that  
$\upp0\h g_{ab}$ is a metric of Minkowski spacetime.
Note that $\upp1\h g_{ab}$ is arbitrary in an {\sc afti-0} spacetime.
That is, eq.(\ref{eq:gab=...}) tells us that 
an {\sc afti-0} spacetime is asymptotically Minkowskian
but how it approaches an asymptotically Minkowski spacetime is not
specified in any way.


\section{Asymptotic Symmetries}\label{sec:symmetry}
In this section, we define asymptotic symmetries 
by considering the transformations that preserve the zero-th order asymptotic structure of 
an {\sc afti-$0$} and an {\sc afftis-$0$} spacetime and analyze their structures.
Basically, this section is a review of the corresponding work of
\cite{AR} but differs in the following points. 
First, the definition of the asymptotic symmetries 
is different due to the difference in the asymptotic structure they are to preserve.
Second, one of the subgroups of the asymptotic symmetry group
is examined in detail for the explicit evaluation of four-momentum 
in the sec.~\ref{sec:conserved}. 
Third, the structure of the asymptotic symmetries 
differ due to the presence of a bounded source at the
late time regime in an {\sc afftis} spacetime.
 
In Minkowski spacetime, the symmetry group, i.e., the Poincar\'e
group is a group of diffeomorphisms $\{ \phi \}$ 
that preserve the structure of the spacetime $\h\eta_{ab}$: 
$\phi^*\h\eta_{ab}=\h\eta_{ab}$. 
Because the structure that is universal to 
{\sc afti-$0$} or {\sc afftis-$0$} spacetimes is 
the zero-th order asymptotic structure, 
we define the asymptotic symmetries of such spacetimes
to be transformations that preserve the asymptotic structures.

First we examine the infinitesimal asymptotic symmetries.
As they are {\em local} transformations, the structure they preserve
is that of an {\sc afti-0} spacetime.

\medskip

\noindent
{\bf DEFINITION:} The {\em infinitesimal Ti symmetry group} 
${\cal L}_{\cal G}$ is a group of equivalence classes
of a $C^\infty$ vector field $\xi^a$ 
that satisfies the following conditions:

(1) $\xi^a$ induces a smooth function $\bm\alpha$ given by
%
\beq
  \bm\alpha := \lim_{\to\bri}\frac{ \Lie_{\xi} \Omega }{ \Omega^2 }\label{eq:Dalp};
\eeq
%
\ \ \ \ \ and 

(2) $\xi^a$ induces a Killing vector of $(\bri,\bm h_{ab})$,
%
\beq
      \Lie_{\bm\xi} \bm h_{ab} \breve= 0.
\eeq
%
Two vector fields $\xi_1^a$ and $\xi_2^a$ are regarded as
equivalent if and only if 
$(\bm\alpha_1,\bm{\xi}_1^a) \breve= (\bm{\alpha}_2,\bm{\xi}_2^a)$.

(Ti stands for
timelike infinity and rhymes with Scri(${\cal I}^{\pm}$) 
as Ashtekar's Spi($i^0$) does.)

\medskip

The first condition demands that $\Psi_\xi^*(\Omega)$
differs from $\Omega$ by terms of $O(\Omega^2)$
so that $\Psi_\xi^*(\Omega)$ also induces a unit hyperboloid
completion and thus $\xi^a$ preserves the zero-th order
asymptotic structure $F\breve=1$, 
where $\Psi$ denotes an infinitesimal map generated by $\xi$.
The second condition ensures that $\xi^a$ preserves the zero-th order
asymptotic structure $\bm q_{ab}\breve=\bm h_{ab}$. 
The equivalence class is defined because the vector fields that belong 
to the same class cannot be distinguished as a generator of
symmetries.

Let us investigate the structure of the infinitesimal Ti 
group. Because its element is a $C^\infty$ vector field, 
${\cal L}_{\cal G}$ has the structure of a Lie algebra. 
The Lie bracket on the pairs is given by
%
\beq
     [(\bm{\alpha},\bm{\xi}^a),
      (\bm{\beta},\bm{\eta}^a)] \hat{=}
        (\Lie_{\bm{\xi}} \bm{\beta}-
         \Lie_{\bm{\eta}}\bm{\alpha}, [\bm\xi,\bm\eta]^a).
                         \label{eq:Lie}
\eeq
%
The introduction of the following subgroup of ${\cal L}_{\cal G}$
helps the investigation of the structure on ${\cal L}_{\cal G}$.

\medskip

\noindent
{\bf DEFINITION:} The {\em infinitesimal Ti supertranslation group} 
${\cal L}_{\cal S}$ is a subgroup of ${\cal L}_{\cal G}$ whose element 
$(\bm{\alpha},\,\bm{\xi})$ satisfies
%
\beq
     \bm{\xi}^a \breve{=} 0.
\eeq
%

\medskip
\noindent

Because the subgroup contains the following infinitesimal Ti {\em translation} group,
we call it infinitesimal Ti {\em supertranslation} group.

\medskip
                    
\noindent
{\bf DEFINITION:} The {\em infinitesimal Ti translation} group 
${\cal L}_{\cal T}$ is a subgroup of ${\cal L}_{\cal S}$ whose element 
$(\bm{\alpha},0)$ satisfies
%
\beq
     \bm D_a\bm D_b\bm{\alpha} - 
              \bm{\alpha} \bm h_{ab}\hat{=} 0
                                  \label{eq:DefAlpha}
\eeq
%
where $\bm D_a$ is the derivative operator associated with the intrinsic metric
$\bm h_{ab}$ on $\bri$.

\noindent
(The group does not depend on the function $\Omega$ used in the
completion of a spacetime because $\xi^a$ vanishes on $\bri$ and 
thus $\bm\alpha$ does not depend on $\Omega$.)

\medskip
\noindent
It is important to note here
that there are {\em four} independent functions that satisfy eq.(\ref{eq:DefAlpha}):
\beq
      \bm{\alpha}_t \hat{=} \cosh\chi,\quad
      \bm{\alpha}_x \hat{=} \sinh \chi \sin \theta \cos \phi,\quad  
      \bm{\alpha}_y \hat{=} \sinh \chi \sin \theta \sin \phi 
            ~~\mbox{and}~~
      \bm{\alpha}_z \hat{=} \sinh \chi \cos \theta \label{eq:alphas}
\eeq
where $(\chi,\theta,\phi)$ is the standard hyperbolic coordinate on
$\bri$.
It can be shown that if $\bri$ is timelike infinity of 
$n$-dimensional spacetime and thus
$\bm h_{ab}$ in eq.(\ref{eq:DefAlpha}) is the $(n-1)$-hyperboloid metric, 
there are $n$ independent functions satisfying the equation.
The number corresponds to the number of possible ``directions'' 
of translation in the spacetime.
When Minkowski spacetime is completed, the translational Killing fields 
correspond to the elements of ${\cal L}_{\cal T}$. 
It will be shown that the group is closely related with 
the asymptotic {\em translation} Killing fields introduced 
in sec.\ref{sec:translation}
and the four-momentum of the {\sc afftis} spacetime introduced 
in sec.\ref{sec:conserved}.

The structure of ${\cal L}_{\cal G}$ can be examined by
noting that ${\cal L}_{\cal S}$ is 
a Lie ideal of ${\cal L}_{\cal G}$~\cite{AR}.
Hence, the quotient ${\cal L}_{\cal G}/{\cal L}_{\cal S}$ has the structure of 
a Lie algebra consisting of cosets $(\{ \bm{\alpha} \},\,\bm{\xi}^a)$
where $\{ \bm{\alpha} \}$ is a set of all the functions smooth on $\bri$. 
Thus, each element of the quotient is characterized by ${\bm\xi}^a$.
In other words,  ${\cal L}_{\cal G}/{\cal L}_{\cal S}$ 
is isomorphic to the group of the Killing fields
of $(\bri,\bm h_{ab})$. 
Recalling that 
such Killing fields is in the Lie algebra of the Lorentz group
${\cal L}_{L}$, we obtain
%
\beq 
     {\cal L}_{\cal G}/{\cal L}_{\cal S}\cong {\cal L}_L
    \qquad\mbox{and thus}\qquad
      {\cal L}_{\cal G} \cong {\cal L}_{\cal S}\stimes {\cal L}_L
\eeq
%
where $\stimes$ denotes the semi-direct sum here.

Next, let us consider the asymptotic symmetry group.
Following the symmetry group of Minkowski spacetime,
we define the asymptotic symmetry group as a group of diffeomorphisms. 
Because {\sc afti-0} spacetimes possess only 
the zero-th order asymptotic structure in common, 
the diffeomorphisms can be only defined on $\bri$. 
However, $\bri$ of an {\sc afti-0} spacetime is not specified enough 
to consider diffeomorphisms generated on itself.
Hence, we consider $\bri$ of
an {\sc afftis-0} spacetimes whose topology is restricted and demand that
$\bri$ be geodesically complete 
so that diffeomorphisms could be generated on it.

\medskip

\noindent
{\bf DEFINITION:} The {\em Ti group} ${\cal G}$ is a group of
diffeomorphisms generated by elements of ${\cal L}_{\cal G}$ 
on $\bri$ that is geodesically complete except for the geodesics
through a point and that has topology of $S^2\times(\real^+-\{ p\})$.

\medskip

Recall that a 3-manifold with negative-constant curvature with
topology of $S^2\times(\real^+-\{ p\})$ is a spacelike 3-hyperboloid
with a removed point. Hence, it is only when 
the removed point is chosen as the origin of the
Lorentz transformations that the elements of ${\cal L}_L$ do generate
diffeomorphisms. Otherwise, the integral curves of the elements of ${\cal L}_L$
would be incomplete due to the point. 
Hence, the Ti group ${\cal G}$ is a semi-direct product of
the infinite dimensional additive group of smooth functions on the
unit spacelike hyperboloid with the Lorentz group around the
removed point, $L_{\removedPt}$:
\beq
     {\cal G} \cong {\cal S} \stimes L_{\removedPt}. \label{eq:group}
\eeq
The above additive group is {\em infinite} dimensional 
in contrast with {\em four}
dimensional additive subgroup of the Poincar\'e group's, the translation group. 
This is because, 
as for transformations of the zero-th order asymptotic structures,
translations are the changes of the function $\Omega$ 
used in completions and the changes are possible 
in ``infinitely different ways''. 
It is interesting
that this structure is very similar to that of the BMS group defined at null
infinity which is a semi-direct product of the infinite dimensional
additive group of {\em conformally weighted functions on the
2-sphere} and the Lorentz group{\cite {BMV,RS,RP}}.

However, it is important to note here 
that the additive group ${\cal S}$ in eq.(\ref{eq:group})
consists of all the possible finite supertranslation. 
This is against our intuition that there are no spacelike translations 
when there exists a source, or a removed point on $\bri$, 
that breaks the symmetries. 
The reason traces back to the fact 
that the Ti group is defined only to preserve the structure of $\bri$
but the structure of $\bri$ is preserved as long as the function $\alpha$
is smooth, regardless of the detailed nature of $\alpha$, which characterizes
supertranslations.
In the next section, we find that a translation
whose associated function is  $\alpha$ transforms the points
in the vicinity of $\bri$ in $\cal F$
by the generator $\Omega D_a \alpha$.
Hence, to take the actions of the supertranslations 
into the symmetry group more appropriately,
we need to define the group ${\cal G}'$ to 
consist of diffeomorphisms on {\em the vicinity} of $\bri$.
Now, the generators of the group must be vanishing in the vicinity of
the removed point on $\bri$, and thus
${\cal G}'$ is a semi-direct product of
the infinite dimensional additive group ${\cal S}_{\removedPt}$ 
of smooth functions
whose gradient vanishes in the vicinity of the removed point
and the Lorentz group around the removed point:
\beq
     {\cal G}' \cong {\cal S}_{\removedPt} \stimes L_{\removedPt}.
\eeq
This modified Ti group ${\cal G}'$ does not contain 
``spacelike supertranslation'' because,
as we shall see in the next section, 
the gradients of the functions $\alpha$ associated with spacelike 
translations do not vanish in the vicinity of the removed point.


\section{asymptotic killing fields}\label{sec:translation}

In this section, we introduce the notion of asymptotic Killing fields.
The introduction helps us to understand that 
the asymptotic symmetries are closely related with the hierarchy 
found in the asymptotic structure and gives a basis for discussing 
how fast a spacetime becomes stationary.

If a vector field $\h\xi^a$ is a Killing vector field, 
the Lie derivative of the metric with respect to $\h\xi^a$ vanishes. 
This fact motivates us to call the vector field $\h\xi^a$
an {\em asymptotic} Killing field if $\Lie_{\h\xi}\h g_{ab}$
{\em asymptotically} vanishes to the limit $\Omega\to0$.

\medskip

\noindent
{\bf DEFINITION:} \enskip A spacetime ($\hat{\cal M}$,$\hat{g}_{ab}$) 
is said to admit an 
{\em asymptotic Killing field} $\h\xi^a$ to order $n$ if 
%
\beq
   \lim_{\to\bri}\Omega^{-n}(\Lie_{\h\xi}\h g)_{\h\mu\h\nu} = 0
\eeq
%
where $(\Lie_{\h\xi}\h g)_{\h\mu\h\nu}:=
\Psi^*(\base{\h e}{\mu}^a\base{\h e}{\nu}^b\Lie_{\h\xi}\h g_{ab})$. 

\medskip


The local asymptotic symmetries defined above are related to the
infinitesimal Ti group defined in the previous section 
as follows.

\medskip

\noindent
{\bf PROPOSITION 1:} \enskip 
In an {\sc afti-$0$} spacetime, 
a vector field $\h\xi^a$ belongs to the infinitesimal Ti group
if and only if $\h\xi^a$ is an asymptotic Killing field
to order 0.
\medskip

\noindent

{\em Proof of If}: The components of the Lie derivative of the metric
with respect to $\h\xi^a$ are given by 
%
\beqn
      \h n^a\,\,\h n^b\, \Lie_{\h\xi}\hat{g}_{ab}  & = &
                              \ \  4\Omega^{-1}\Lie_{\h\xi}\Omega 
                            - 2F^{-1}\Lie_{[\h\xi,n]}\Omega 
                            - F^{-1}\Lie_{\h\xi}F           \nonumber \\ 
     \base{\hat e}{I}^a\,\h n^{b}\, \Lie_{\h\xi}\hat{g}_{ab}  & = &
               \left[ \,\Omega^{-1} F^{-1/2}D_{m}( \Lie_{\h\xi}\Omega ) 
                - \Omega q_{ma} F^{-1/2}\Lie_{\h\xi}n^{a} 
               \right] \base{e}{I}^m                      \label{eq:LgabRAW}\\
     \base{\hat e}{I}^a \base{\hat e}{J}^b \Lie_{\h\xi}\hat{g}_{ab} & = & 
                \left[ - 2 \Omega^{-1}(\Lie_{\h\xi}\Omega) q_{mn}
                +{q_{m}}^{a}{q_{n}}^{b}\Lie_{\h\xi}q_{ab} 
                \right] \base{e}{I}^m \base{e}{J}^n      \nonumber
\eeqn
%
in terms of smooth functions and tensors.
Because $\h\xi^a$ is an asymptotic Killing field to order 0,
all the components must be of $O(\Omega^1)$ by the definition. 
Hence, from the time-time component of $\Lie_{\h\xi}\h g_{ab}$ 
in the above equation, we find that 
$\alpha:=\Lie_{\h\xi}\Omega/\Omega^2$ must be a smooth
function. Hence, $\bm\alpha:=\lim_{\to\bri}\alpha$ is 
a smooth function on $\bri$ and the above identities become 
\beqn
      \h n^a\,\,\h n^b\, \Lie_{\h\xi}\hat{g}_{ab}  & = &
            \ \, \qquad~ 0 \qquad~ +\Omega{\cal L}_{\h\xi}{}^{(1)}F + O(\Omega^2)
                                      \nonumber \\ 
     \base{\hat e}{I}^a\,\h n^{b}\, \Lie_{\h\xi}\hat{g}_{ab}  & = &
            \left[\,
                \qquad~ 0 \qquad~
                 + \Omega (\,D_{m}\alpha - q_{ma}{\cal L}_{\h\xi}n^{a} \,)
                 + O(\Omega^2) \,
            \right] \base{e}{I}^m     \label{eq:LgabALPHA}\\
     \base{\hat e}{I}^a \base{\hat e}{J}^b \Lie_{\h\xi}\hat{g}_{ab} & = & 
                \left[\,
                  {q_{m}}^{a}{q_{n}}^{b}{\cal L}_{\h\xi}q_{ab}
                     - 2\Omega\alpha q_{mn}\, 
                \right] \base{e}{I}^m \base{e}{J}^n          \nonumber
\eeqn
%
in {\sc afti-$0$} spacetimes where eq.(\ref{eq:unitF}) is used in the
evaluation. Next, from the space-space component, we find 
%
\beq
        {q_{m}}^{a}{q_{n}}^{b}\Lie_{\h\xi}q_{ab} \breve= 0 
\eeq
%
must be satisfied so that all the components of $\Lie_{\h\xi}\h g_{ab}$
are of $O(\Omega^1)$. The equation above induces
%
\beq
     \Lie_{\h{\bm\xi}} \bm h_{ab} \breve= 0.             
\eeq
%
on $\bri$. The equation tells us that 
$\h\xi^a$ induces a Killing field of $(\bri,\bm h_{ab})$.

To summarize, an asymptotic Killing field to order 0 
generates a smooth function $\bm\alpha$ on $\bri$ and induces a Killing field of
$(\bri,\bm h_{ab})$.  
The vector field belongs to the infinitesimal Ti group.~$\Box$

\medskip
\noindent
{\em Proof of Only If:} \enskip If $\h\xi^a$ belongs to the 
infinitesimal Ti group,
$\alpha:=\Lie_\h\xi\Omega/\Omega^2$ is a smooth function
from the definition. Thus, in {\sc afti} spacetimes 
eqs. (\ref{eq:LgabALPHA}) are satisfied. 
Further, $\h\xi^a$ induces a Killing field of $(\bri,\bm h_{ab})$.
Hence, the first term in the space-space component vanishes
resulting in $(\Lie_{\h\xi}\h g)_{\h\mu\h\nu}=O(\Omega^1)$.
That is, $\h\xi^a$ is an asymptotic Killing field to order 0.~$\Box$

\medskip

Eqs.(\ref{eq:LgabALPHA}) show that 
it was appropriate to define the zero-th order asymptotic structure
such that it does not contain the equivalence class of $\Omega$
defined by eq.(\ref{eq:eqvclass}).
As we have seen above, if the components of $\Lie_{\h\xi}\h g_{ab}$
are of order $O(\Omega)$, then the vector field $\xi^a$ 
preserves the zero-th order asymptotic structure.
However, the vector field $\xi^a$ preserves 
the equivalence class of $\Omega$,
only when $D_{m}\alpha - q_{ma}\Lie_{\h\xi}n^{a}\breve=0$ were satisfied in addition,
or the time-space components of $\Lie_{\h\xi}\h g_{ab}$ were of $O(\Omega^2)$,
as it can be seen from eqs.(\ref{eq:LgabALPHA}).
(For the details on the preservation of the class, see \cite{AR}.) 
The fact that the time-space components of $\Lie_{\h\xi}\h g_{ab}$
must be $O(\Omega^2)$ instead of $O(\Omega^1)$
to preserve the class, 
indicates that the equivalence class of $\Omega$
is of {\em an order higher} asymptotic structure.

Next, we investigate the asymptotic translational Killing fields. 
This is to see how fast an {\sc afti-0} spacetime becomes stationary.

\medskip

\noindent
{\bf PROPOSITION 2:} \enskip 
In an {\sc afti-0} spacetime, 
%
\beqn
     \h\xi_t{}^a &=&  \Omega \left[\, \sinh\chi\base{e}{1}^a \,
                           \right] 
                     +\Omega^2 F^{-1/2}\cosh\chi\ n^a \nonumber\\
     \h\xi_x{}^a &=&  \Omega \left[\, \cosh\chi\sin\theta\cos\phi\base{e}{1}^a
                                   +\cos\theta\cos\phi\base{e}{2}^a
                                   -\sin\phi\base{e}{3}^a \,
                           \right]
                     +\Omega^2 F^{-1/2}\sinh\chi\ n^a \nonumber\\
     \h\xi_y{}^a &=&  \Omega \left[\, \cosh\chi\sin\theta\sin\phi\base{e}{1}^a
                                   +\cos\theta\sin\phi\base{e}{2}^a
                                   +\cos\phi\base{e}{3}^a \,
                           \right]
                     +\Omega^2 F^{-1/2}\sinh\chi\ n^a \nonumber\\
     \h\xi_z{}^a  &=& \Omega \left[\, \cosh\chi\cos\theta\base{e}{1}^a
                                   -\sin\theta\base{e}{2}^a \,
                           \right] 
                     +\Omega^2 F^{-1/2}\sinh\chi\ n^a, \label{eq:xiT}
\eeqn
%
are admitted as asymptotic Killing fields to order one, where 
$e_{\ssr 1} = \partial_\chi,
~e_{\ssr 2} = (\sinh^{\ssr -1}\!\chi) \partial_\theta$ and
$e_{\ssr 3} =  (\sinh^{\ssr -1}\!\chi\sin^{\ssr -1}\!\theta )\partial_\phi$.
$\{ \h\xi_{x^\mu}\}_{\s{\mu=0,1,2,3}}$ 
$(\{ {x^\mu} \}=\{ t,x,y,z \})$ 
are four {\em independent asymptotic translational Killing fields}
in the sense that, with respect to the physical metric $\h g_{ab}$,
their norms are constant and they are orthogonal to each other:
%
\[
     \hat{g}_{ab}\,(\h\xi_{x^\mu})^a\, (\h\xi_{x^\nu})^b = \mp\delta_{\mu,\nu}
\]
%
where $-$ for $\mu=0$ and $+$ for $\mu=1,2,3$.

\medskip
\noindent
{\em Proof:}\enskip
The space components of the above asymptotic translational Killing
vector fields are of $O(\Omega)$ and the time components are of $O(\Omega^2)$.
Thus, the expansion of the vector field $\h\xi^a$ is given by 
%
\beq
     \h\xi^a = 
         \sum_{\ell=1}^\infty \Omega^\ell\up{(\ell)}\xi^{\ssr I}  \base{e}{I}^a   
        +\sum_{\ell=2}^\infty \Omega^\ell\up{(\ell)}\xi^{\ssr 0} (\ppp_\Omega)^a
                  \label{eq:xi...}
\eeq 
%
where $\{ \up{(n)}\xi^\mu \}$ is a set of smooth functions
independent of $\Omega$. 
Hence, 
%
\beq
     \alpha := \Lie_{\h\xi}\Omega / \Omega^2 
              = \sum_{\ell=2}^\infty\up{(\ell)}\xi^{\ssr 0} \Omega^{\ell-2}
\eeq 
%
is a smooth function. Thus, the components of $\Lie_{\h\xi}\h g_{ab}$
are given by
%
\beqn
      \h n^a\, \,\h n^b\,\Lie_{\h\xi}\hat{g}_{ab}  &=& 
                \,\ \  \qquad\qquad\quad 0 \quad\qquad\qquad 
                +O(\Omega^2)                       \nonumber \\ 
      \base{\hat e}{I}^a \,\h n^{b}\,\Lie_{\h\xi}\hat{g}_{ab}  &=& 
            \left[\,\, \ \quad 
                \Omega( D_{m}\alpha - \1\xi^{\ssr J}\base{e}{J}_m ) \quad\ 
                 + O(\Omega^2) \,
            \right] \base{e}{I}^m      \label{eq:L(1)xigab}\\
      \base{\hat e}{I}^a \base{\hat e}{J}^b 
                 \Lie_{\h\xi}\hat{g}_{ab}   &=&   
            \left[\,
                \Omega({q_{m}}^{a}{q_{n}}^{b}
                           \Lie_{\1\xi^{\ssr J}\base{e}{J}}q_{ab} - 2 \alpha q_{mn})
                + O(\Omega^2)\,
            \right] \base{e}{I}^m \base{e}{J}^n.
                                         \nonumber
\eeqn
%
From the equations above, we see that $(\Lie_\h\xi \h g)_{\h\mu\h\nu}=O(\Omega^2)$
if the space components of the $O(\Omega^1)$ terms of the vector field
$\h\xi^a$ satisfy
%
\beqn
    D_{m}\alpha -\1\xi^{\ssr J}\base{e}{J}_m\breve= 0 \qquad \mbox{and} \qquad
    {q_{m}}^{a}{q_{n}}^{b}\Lie_{\1\xi^{\ssr J}\base{e}{J}}q_{ab} - 2\alpha q_{mn}
                            \breve= 0.  \label{eq:ORGcond}
\eeqn
%
These equations are tangential to $\bri$ and thus 
induce 
%
\beq
   \1\bm{\xi}^a\, \breve= \, \bm D^a\bm\alpha 
       \qquad\mbox{and}\qquad
   \Lie_{\1\bm\xi}\bm h_{ab} - 2{\bm\alpha}\bm h_{ab} \breve= 0
\eeq
%
on $\bri$ without any loss of information
where $\1\bm\xi^a$ and $\bm\alpha$ denote the vector field and 
the function induced by
$\1\xi^{\ssr J}\base{e}{J}^a$ and
$\lim_{\to\bri}\alpha:=\up{(2)}\xi^{\ssr 0}$ on $\bri$ respectively.
Thus, the conditions can be restated as
%
\beqn
       &\bm D_a\bm D_b\,\bm\alpha 
            -\bm\alpha\,\bm h_{ab}\breve= 0 &  \label{eq:DDa-ah}\\
       &\1\bm{\xi}^a\, \breve= \, \bm D^a\bm\alpha.&  \label{eq:xi-a}
\eeqn
%
In other words, $(\Lie_\h\xi \h g)_{\h\mu\h\nu}=O(\Omega^2)$
is satisfied if the leading term of the time component of $\h\xi$,
$\bm\alpha=\up{(2)}\bm\xi^{\ssr 0}$, satisfies
eq.(\ref{eq:DDa-ah}) and if those of the space components $\1\bm\xi^a$
are given by
eq.(\ref{eq:xi-a}). Note here that eq.(\ref{eq:DDa-ah}) is the same 
as eq.(\ref{eq:DefAlpha}) used in defining the infinitesimal 
Ti {\em translation}
and that $\{ \bm\alpha_{x^\mu} \}$ in eq.(\ref{eq:alphas})
are the four independent smooth functions that satisfy the equation.
Then, from  eq.(\ref{eq:xi-a}), we see that 
the corresponding independent space components are
%
\beqn
  \1\bm{\xi}_t{}^a\ &\breve=&\ \sinh\chi\base{\bm e}{1}^a 
                                                \nonumber\\
  \1\bm{\xi}_x{}^a\ &\breve=&\ \cosh\chi\sin\theta\cos\phi\base{\bm e}{1}^a
                                  +\cos\theta\cos\phi\base{\bm e}{2}^a
                                  -\sin\phi\base{\bm e}{3}^a   \nonumber\\
  \1\bm{\xi}_y{}^a\ &\breve=&\ \cosh\chi\sin\theta\sin\phi\base{\bm e}{1}
                                  +\cos\theta\sin\phi\base{\bm e}{2}^a
                                  +\cos\phi\base{\bm e}{3}^a   \nonumber\\
  \1\bm{\xi}_z{}^a\ &\breve=&\ \cosh\chi\cos\theta\base{\bm e}{1}^a
                                 -\sin\theta\base{\bm e}{2}^a. 
                        \label{eq:(1)xiT}
\eeqn
%
The space components of the vector fields in eqs.(\ref{eq:xiT}) do take this
form and thus are asymptotic Killing fields to order one in $\Omega$.~$\Box$.

\medskip

As a corollary of  proposition 2, 
we obtain the relation between the infinitesimal 
Ti translational group ${\cal L}_{\cal T}$ and 
the above asymptotic translational Killing fields.

\medskip

\noindent
{\bf COROLLARY:} The asymptotic translational Killing fields to order
one, eqs.(\ref{eq:xiT}), belong to the infinitesimal
Ti translational group ${\cal L}_{\cal T}$.
 
\medskip
\noindent
{\em Proof:} \enskip Admitted as asymptotic Killing fields 
to  order one, the asymptotic translational Killing fields belong to the
infinitesimal Ti group according to proposition 1. 
In addition, the fields vanish on $\bri$ and generate 
smooth functions satisfying eq.(\ref{eq:DefAlpha}).~$\Box$

\medskip

Let us consider the finite translations
in {\sc afftis-0} spacetimes. 
From eq.({\ref{eq:xi...}) and ({\ref{eq:xi-a}), 
we see that the generators of translations
vanish on $\bri$ but transform the points in vicinity of $\bri$
in the direction of $\Omega D^a\alpha$, or $\1\xi^a$. 
Although the generator $\1\xi_t{}^a$ of a timelike translation
vanishes in the vicinity of the removed point $\chi=0$ on $\bri$,
the generators $\1\xi_x{}^a,\1\xi_y{}^a$ and $\1\xi_z{}^a$ of
spacelike translations do not. 
Hence, there exist only timelike translation symmetry $\h\xi_t$ and
no spacelike translation symmetries in {\sc afftis-0} spacetimes.
In other words, the existence of source in an {\sc afftis-0}
spacetime breaks the spacelike translation symmetries but leaves the
timelike symmetries.

From the above discussion, it can be said that 
if the energy-momentum tensor of a spacetime falls off at a rate faster than 
$\Omega^2$, the spacetime becomes asymptotically 
stationary at the rate of $\Omega^2$.

We conclude the section with a remark on the relation of the
asymptotic symmetries with the hierarchy in asymptotic structure. 
We have shown that asymptotic translational 
Killing fields are admitted to order 1 in {\sc afti-0} spacetimes.
However, we can see that the fields are not admitted to order 2 in general.
For the vector fields to be admitted to order 2, the coefficients of the
$O(\Omega^2)$ terms in eqs.(\ref{eq:L(1)xigab}) need to vanish. These terms 
contain the higher order asymptotic structure, such as $\1 F$.
For example, the $O(\Omega^2)$ term of the time-time component is 
$\Omega^2[D^a\1 FD_a\alpha+\1 F\alpha-2 n^a\nabla_a\alpha]$.
These higher order asymptotic quantities are {\em arbitrary}
in {\sc afti-0} spacetimes. Hence, in general, the $O(\Omega^2)$ terms
in eqs.(\ref{eq:L(1)xigab}) do not vanish 
and thus asymptotic translational Killing fields to order two 
are not admitted in general {\sc afti-0} spacetimes.
The asymptotic translational Killing fields to order 2 are
the asymptotic symmetries that belong to
the zero-th order asymptotic structure.

\section{gravitational fields}\label{sec:fields}

In this section, we investigate the gravitational fields
and find hierarchy in their asymptotic behaviors.

Because the Ricci tensor of $\h g_{ab}$ is reserved for the definition 
of {\sc afti-$n$} spacetimes, we examine the Weyl tensor $\h C_{abcd}$ to
investigate the gravitational fields.
For convenience of the discussion, 
decompose the Weyl tensor into the electric part 
$\h E_{ab}:=\h C_{ambn}\h n^m \h n^n$ and 
the magnetic part $\h B_{ab}:={}^* \h C_{ambn}\h n^m \h n^n$
where ${}^* \h C_{ambn}$ denotes the dual of the 2-form $\h C_{\s[am\s]bn}$.
In terms of quantities associated with the $\Omega$-const$.$
surfaces and the physical components of the energy-momentum tensor, 
$\h E_{ab}$ and $\h B_{ab}$ are given by
%
\beqn
      \hat{E}_{ab} & = & \h K_{ar}\h K^r{}_b -\Lie_{\h n}\h K_{ab} 
        + \h D_{\s(a} \h a_{b\s)} + \h a_a \h a_b
        + \frac12(\h q^r_a \h q^s_b - \h q_{ab}\h n^r \h n^s)\h L_{rs}\nnb\\
      \hat{B}_{ab} & = & \h\EEE_{ra}{}^s\h D^r\h K_{bs} 
                        + \frac12\h\EEE_{ab}{}^r\h n^s\h L_{rs}
\eeqn
%
where $\h L_{ab}:=\h R_{ab}-\frac16\h R\h g_{ab}$ and 
$\h a_a:=\h q_{ar}\h n^s\V_s \h n^r$.

In {\sc afti-0} spacetimes, we find that 
$\h E_{ab}$ and $\h B_{ab}$ vanish on $\bri$:
%
\beq
    \h E_{ab} = \sum_{\ell=1}^{\infty}\Omega^\ell \up{(\ell)} E_{ab}
       \hspace{5ex} \mbox{and}  \hspace{5ex}
    \h B_{ab} = \sum_{\ell=1}^{\infty}\Omega^\ell \up{(\ell)} B_{ab}
                                           \label{eq:^E=,^B=}
\eeq
where $\up{(n)} E_{ab}$ and $\up{(n)} B_{ab}$ are defined by
%
\beqn
     \up{(1)} E_{ab}:= \lim_{\to\bri}\Omega^{-1} \h E_{ab}, \hspace{16ex}
     \up{(n)} E_{ab}:= \lim_{\to\bri}\Omega^{-n} 
                        (\h E_{ab}-\sum_{\ell=1}^{n-1} \up{(\ell)} E_{ab})
                                  \hspace{3ex}\mbox{for $n\geq2$}, \\
     \up{(1)} B_{ab}:= \lim_{\to\bri}\Omega^{-1} \h B_{ab} 
          \hspace{7ex} \mbox{and}  \hspace{7ex}
     \up{(n)} B_{ab}:= \lim_{\to\bri}\Omega^{-n} 
                        (\h B_{ab}-\sum_{\ell=1}^{n-1} \up{(\ell)} B_{ab})
                                  \hspace{3ex}\mbox{for $n\geq2$}.
\eeqn
%
Note here that, although they depend on the choice of $\Omega$,
$\up{(n)} E_{ab}$ and $\up{(n)} B_{ab}$ are defined,
regardless of whether their lower order terms vanish. This is in contrast
with the treatment of \cite{AR} where the subsequent order terms are defined 
only when their lower order terms vanish. Their leading terms are given by
%
\beqn
   \1 E_{ab} &=& \frac12 ( D_a D_b \!-\! h_{ab} )\1 F 
                + \frac12 \base{e}{I}_{a}\base{e}{J}_b 
                   ( \up{(3)}\h L_{\ssr{\h I\h J}}
                     +\delta_{\ssr{I,J}}\!\!\up{(3)}\h L_{\ssr{\h0\h0}}) \\
   \1 B_{ab} &=& \frac12\,\EEE_{ra}{}^s D^r \1 K_{bs}
                + \frac12\,\EEE_{ab}{}^r \base{e}{I}_r
                  \up{(3)}\h L_{\ssr{\h 0\h I}}
\eeqn
%
where 
   $\1 K_{ab} := \lim_{\to\bri} \Omega^{-1}(K_{ab}-h_{ab})$,
   $\h L_{\ssr{\h0\h0}}:=\Psi^*[\h L_{ab} \h n^a\h n^b]$,
   $\h L_{\ssr{\h0\h I}}:=\Psi^*[\h L_{ab} \h n^{\s(a}\base{\h e}{I}^{b\s)}]$,
   $\h L_{\ssr{\h I\h J}}:=\Psi^*[\h L_{ab} \base{\h e}{I}^a\base{\h e}{J}^b]$
and $\up{(n)}\h L_{\h\mu\h\nu}$ denotes the coefficient of the 
$O(\Omega^n)$ term of $\h L_{\h\mu\h\nu}$.
$\1 E_{ab}$ and $\1 B_{ab}$ depend on 
$\up{(3)}\h L_{\h\mu\h\nu}$ that consists of the higher
order asymptotic structure such as $\1 F$ and $\1 K_{ab}$.
Because these quantities are arbitrary in {\sc afti-0} spacetimes,
we cannot conclude anything more than 
that $\h E_{ab}$ and $\h B_{ab}$ vanish on $\bri$.

Let us examine the gravitational fields in {\sc afti-1} spacetimes.
The fall-off condition on the energy-momentum tensor demands 
$\up{(3)}\h L_{\h\mu\h\nu}$ vanish on $\bri$ and we obtain
%
\beq
   \1 E_{ab} \breve= \frac12 ( D_a D_b \!-\! h_{ab} )\1 F  
       \hspace{5ex} \mbox{and}  \hspace{5ex}
   \1 B_{ab} \breve= \frac12\,\EEE_{ra}{}^s D^r \1 K_{bs}. \label{eq:1Bab=}
\eeq
%
The equation on $\1 E_{ab}$ induces 
%
\beq
   \1\bm E_{ab} \breve= \frac12 ( \bm D_a \bm D_b \!-\! \bm h_{ab})\1\bm F  
       \label{eq:1Eab=}
\eeq
%
on $\bri$. Using eq.(\ref{eq:constCRV}), 
it can be shown that $\1\bm E_{ab}$ obeys the field equation 
%
\beq
   \bm D_{\s[a} \1\bm E_{b\s]c} \breve= 0. \label{eq:1EabFIELDeq}
\eeq
%
Contracting over $a$ and $b$ and recalling that $\1\bm E_{ab}$ is
traceless by definition, 
we find that $\1\bm E_{ab}$ is divergence-free:
%
\beq
   \bm D_a \1\bm E^{ab} \breve= 0.
\eeq
%
It can be shown that $\1\bm B_{ab}$ satisfies the same field equation. 
Contracting the Bianchi identity $\V_{\s[a} \h R_{bc\s]}{}^{de}=0$ 
with $\h n^a\h\EEE^{bc}{}_m\h\EEE_{den}$, we obtain
%
\beqn
         \h\EEE^{ac}{}_m \h D_a \h B_{cn} &-& \h n^a \h \V_a \h E_{mn} 
      +  \h\EEE^{ac}{}_m \h\EEE^{de}{}_n \h K_{ad}\h E_{ce}
                                      \nonumber\\
     &-& 2\h\EEE^{bc}{}_{(m}\h B_{n)c} \h a_b 
      +  \frac12 \left( \h n^{[a}\h q^{c]e} \h q_{mn}
                       - \h n^{[a} {\h q^{c]}}_n {\h q^e}_m 
                  \right) 
                      \h \V_a \hat{L}_{ce} = 0 \label{eq:VRab}
\eeqn
%
from which we find, in {\sc afti-$n$} spacetimes,
%
\beq
     D_{\s[a}\1 B_{b\s]c} = O(\Omega) + O(\Omega^n)
\eeq
%
where we have used the equations
$n^r\V_r\base{\h e}{I}_a=-\base{e}{I}_a+O(\Omega)$,
$\Omega\cdot n^r\V_r\h L_{\ssr{\h I\h J}}=O(\Omega^{n+3})$,
$\h L_{\ssr{\h\mu\h\nu}}=O(\Omega^{n+3})$ and 
$D_a\h L_{\ssr{\h 0\h I}}=O(\Omega^{n+3})$ to evaluate the derivative 
of $\h L_{ab}$:
%
\beqn
       \left( \h n^{[a}\h q^{c]e} \h q_{mn}
                       - \h n^{[a} {\h q^{c]}}_n {\h q^e}_m 
                  \right) 
                      \h \V_a \hat{L}_{ce} 
       &=& \Omega^{-1}\!\left[\,
                q_{mn}\left(
                \Omega n^a\V_a\h L_{\ssr{\h I\h J}}-2\h L_{\ssr{\h I\h I}}
                +2\h L_{\ssr{\h0\h0}} -\base{e}{I}^e D_e \h L_{\ssr{\h0\h I}}
                      \right)\right.    \nnb\\ 
       & &  \qquad\left. +\base{e}{I}_m\base{e}{J}_n
                (\Omega n^a\V_a\h L_{\ssr{\h I\h J}}-\h L_{\ssr{\h I\h J}})
                     +\base{e}{I}_m D_n \h L_{\ssr{\h0\h I}} 
                      + O(\Omega^{n+4})
                     \right]\nnb\\
       &=& O(\Omega^{n+2}).\nnb
\eeqn
%
Hence, in {\sc afti-1} spacetimes, the equation $D_{\s[a}\1 B_{b\s]c}\breve=0$
is satisfied and induces
%
\beq
   \bm D_{\s[a}\1\bm B_{b\s]c} \breve = 0  \label{eq:1BabFIELDeq}
\eeq
%
on $\bri$. Thus $\1\bm B_{ab}$ is also divergence-free:
%
\beq
   \bm D_a \1\bm B^{ab}\breve = 0.
\eeq
%
Here, it is important to note that the explicit form of the function $\1\bm F$ can be
obtained. Because $\1\bm E_{ab}$ is traceless by definition,
the contraction of eq.(\ref{eq:1Eab=}) gives a differential
equation for the function $\1\bm F$:
%
\beq
    \bm D^a \bm D_a \1 \bm F -  3 \1 \bm F \breve= 0.\label{eq:diff1F}
\eeq
%
We have obtained the non-trivial equation
because the Einstein equation is used in a non-trivial way 
to obtain eq.(\ref{eq:1Eab=}). 
That is, the terms that are not traceless,
$\frac12\base{e}{I}_{a}\base{e}{J}_b( \up{(3)}\h L_{\ssr{\h I\h J}}
+\delta_{\ssr{I,J}}\!\!\up{(3)}\h L_{\ssr{\h0\h0}})$, are discarded to
obtain the equation. In contrast,
we do not obtain any non-trivial equations from noting
that $\1 B_{ab}$ is traceless because the discarded terms 
to obtain the corresponding equation (\ref{eq:1Bab=})are traceless.
The general solution of eq.(\ref{eq:diff1F}) takes the form 
%
\beq
    \1 \bm F (\chi,\theta,\phi)=
            \sum_{ {\ell\geq 0}\atop{|m|\leq\ell} }
                   \bm a_{\ell m} \1 \bm F_{\ell m}(\chi,\theta,\phi)
                    \label{eq:(1)F}
\eeq
%
where 
%
\beq
    \1 \bm F_{\ell m} (\bm\chi,\bm\theta,\bm\phi):=
        \frac1{ \sqrt{\sinh\bm\chi} }\bm P^{\ell+\frac12}_{\frac32}(\cosh\bm\chi) 
                    \bm Y_{\ell m}(\bm\theta,\bm\phi).
\eeq
%
Here, $P^{\mu}_{\nu}(z)$ denote the associated Legendre functions defined by
%
\beqn 
      P^{\mu}_{\nu}(z):= \frac{1}{ \Gamma(1-\mu) }
                  \lmk \frac{z+1}{z-1} \rmk^{\mu/2} 
                     F \lmk -\nu,\nu+1,1-\mu; \frac{1-z}{2} \rmk 
      \quad \mbox{for} \quad z>1  
\eeqn
%
where $F(\alpha,\beta,\gamma; z)$ are the hypergeometric functions.
$Y_{\ell m}$ stands for the spherical harmonics with 
their coefficients chosen as
%
\beq
     \bm Y_{\ell m}(\theta,\phi) = 
       \cases{ & $\sqrt{2\pi}\ \bm P^{m}_{\ell}(\cos\bm\theta)\cdot\cos m\bm\phi
                         \qquad m \geq 0$ \cr
               & $\sqrt{2\pi}\ \bm P^{m}_{\ell}(\cos\bm\theta)\cdot\sin m\bm\phi
                         \qquad m<0$ \cr }
\eeq
%
so that their physical meanings can be conveniently read in the next section.
Explicitly, we obtain
%
\beqn
 \1 \bm F_{00} &=& 2\left( 2\sinh\bm\chi + \frac{1}{\sinh\bm\chi} \right) 
                                                          \label{eq:1F00}\\ 
 \1 \bm F_{11} &=& 2\left( 2\cosh\bm\chi - \frac{\cosh\bm\chi}{\sinh^2\bm\chi} \right)
                          \sin\bm\theta \cos\bm\phi.
\eeqn
%
When the Schwarzschild spacetimes are
completed, the function $F$ indeed consists of the above functions.
See the appendix for the explicit calculations.

To define an angular momentum, it is important to see 
the conditions for  $\2\bm B_{ab}$ to be  divergence-free.
This can be done using the Bianchi identity $\V_{\s[a} \h R_{bc\s]}{}^{de}=0$ 
contracted with $\h\EEE^{abc}\h n_d \h q^s{}_e$,
%
\beq
   \h D_r \h B^{rs} = \frac12 \h\EEE_s{}^{rp}\h n^q\V_r \h L_{pq}.
\eeq
%
Keeping in mind that $\1\bm B_{ab}$ is divergence-free, 
we find that the above equation induces 
%
\beq
     \bm D_r \up{(2)}\bm B^{rs}\breve= \up{(4)}\bm L_{\ssr{IJ}}\bm\EEE^{srp}
                                   \base{\bm e}{I}_r\base{\bm e}{J}_p.
\eeq
%
Hence,  $\2\bm B_{ab}$ is not divergence-free in general {\sc afti-1} 
spacetimes. 
However, in {\sc afti-2} 
spacetimes $\up{(4)}\bm L_{\ssr{IJ}}$ vanishes on $\bri$ and 
$\2\bm B_{ab}$ is divergence-free:
%
\beq
     \bm D_r \up{(2)}\bm B^{rs}\breve= 0.
\eeq
%
Note here that $\1 B_{ab}$ does not need to vanish on $\bri$ 
for $\up{(2)}\bm B^{ab}$ to be divergence-free.
This point is unclear in \cite{AR}.

We conclude the section with a remark on the relation of the
asymptotic behavior of gravitational fields 
with the hierarchy in the asymptotic structure. 
The gravitational fields of each order, 
$\up{(n)} E_{ab}$ and $\up{(n)} B_{ab}$,
are constructed from the asymptotic structure of the corresponding order.
Hence, if the energy-momentum tensor falls off at a low rate and
the spacetime lacks the higher order asymptotic structure, the
asymptotic behaviors of the higher order gravitational fields are 
arbitrary. In an {\sc afti-0} spacetime,
$\up{(0)} E_{ab}$ and $\up{(0)} B_{ab}$ vanish on $\bri$
and $\up{(\ell)} E_{ab}$ and $\up{(\ell)} B_{ab}$
take arbitrary forms for $\ell\geq 1$.
In an {\sc afti-1} spacetime, $\up{(1)} E_{ab}$ and $\up{(1)} B_{ab}$ 
induces divergence-free tensor fields on $\bri$ that satisfy the field
equations (\ref{eq:1EabFIELDeq})(\ref{eq:1BabFIELDeq}) respectively,
and $\up{(1)} E_{ab}$ takes the form given by
eqs.(\ref{eq:(1)F})(\ref{eq:1Bab=}). $\up{(\ell)} E_{ab}$ and 
$\up{(\ell)} B_{ab}$ take arbitrary forms for $\ell\geq2$.
In an {\sc afti-2} spacetime, $\up{(2)} B_{ab}$ induce 
a divergence-free tensor field on $\bri$.


\section{conserved quantities}\label{sec:conserved}

In this section, we define two conserved quantities,
namely four-momentum of {\sc ati-$1$} spacetimes and 
angular momentum of {\sc ati-$2$} spacetimes, which 
are associated with the asymptotic symmetries that the spacetimes possess. 

As Minkowski spacetime admits infinitesimal {\em translation} symmetries,
it possesses a conserved four-momentum. 
The four-momentum is defined as a linear mapping 
from the four-dimensional vector space of 
infinitesimal translations to real numbers.
Because {\sc ati-$n$} spacetimes possess infinitesimal Ti
translations, we expect that four-momentum of {\sc ati-$n$}
spacetimes can be also defined
as a linear mapping of infinitesimal Ti translations to real numbers.

The four-momentum representing the state of the spacetime
should be defined with information from the gravitational fields.
However, {\sc ati-$0$} spacetimes possess no asymptotic gravitational fields,
as we have seen in the previous section.
This means that four-momentum cannot be defined in {\sc ati-$0$} spacetimes.
Now, consider {\sc ati-$1$} spacetimes.
Here, we have gravitational fields $\1\bm E_{ab}$ and $\1\bm B_{ab}$.
On the other hand, infinitesimal Ti translations may be represented 
by $\bm D^a\bm\alpha_{\cal T}$, where $\alpha_{\cal T}$ is a smooth function
generated by the elements of ${\cal L}_{\cal T}$.
The vector field is a conformal Killing field of 
$(\bri,\bm h_{ab})$ by definition:
\beq
     \bm D_{\s(a} \bm D_{b\s)}\bm\alpha_{\cal T}\breve= 
              2\bm\alpha_{\cal T}\bm h_{ab}.
\eeq
With these $\1\bm E_{ab}$ and $\bm D^a\bm\alpha_{\cal T}$,
the four-momentum of an {\sc ati-$1$} spacetime can be defined as follows.

\medskip
\noindent
{\bf DEFINITION:} Four-momentum $\bm P$ of an {\sc ati-$1$} spacetime
with a region $\cal A$ on $\bri$ that satisfies

1) $\cal A$ is bounded by cross sections ${\cal S}_a$ and ${\cal S}_b$ 
that do not intersect with each other, i.e., 
$\ppp {\cal A} = {\cal S}_a \cap {\cal S}_b$ and 
${\cal S}_a \cap {\cal S}_b=\emptyset$,

2) $q_{ab}$ and $n^a$ have smooth limit to $\cal A\subset\bri$ 
and functions $\bm \alpha_{\cal T}$ generated by elements of 
${\cal L}_{\cal T}$ are smooth on $\cal A$

\noindent
is a linear mapping of $\bm\alpha_{\cal T}$
to real numbers defined by
\beq
    \bm P: \bm\alpha_{\cal T} \mapsto\real 
    \quad  \stackrel{\rm def}{\Leftrightarrow} \quad
     \bm P(\bm\alpha_{\cal T}) := - \frac1{8\pi}\int_{\cal S} 
         {\bm\EEE_{ab}}^m \1\bm E_{mn} \bm D^n\bm\alpha_{\cal T}   
                       \label{eq:defP}
\eeq
where $\cal S$ is a cross section of the region $\cal A$.

\medskip

\noindent

Note that, in general, {\sc ati-$1$} spacetimes do not 
possess a region $\cal A$ satisfying the above properties;
$q_{ab}$ and $n^a$ are demanded to have smooth limits to $\bri\cap{\cal F}$
in {\sc ati-$1$} spacetimes
but the region $\bri\cap{\cal F}$ does not need to satisfy the above properties 1) and 2).
However, we need such a region on $\bri$ to define a ``conserved quantity''
as we shall see below.

The four-momentum can be considered as a ``conserved quantity'' 
because it is invariant under a change of the cross
section ${\cal S}$ chosen for the evaluation.
This is because the 2-form integrand is closed in $\cal A$: the 3-form 
$\bm D_{[c}{\bm{\epsilon}_{\,ab]}}^m \1 \bm E_{mn} {\bm D}^n 
{\bm\alpha}_{\cal T}$ is proportional to 
%
\beqn
      \bm\epsilon^{cab}
            \bm D_c \left( 
        {\bm\epsilon_{ab}}^m\1 \bm E_{mn} \bm D^n \bm\alpha_{\cal T} 
                   \right)  &\breve=&
      ( \bm D^m \1 \bm E_{mn} ) \bm D^n \bm\alpha_{\cal T} 
        + \1 \bm E_{mn} (\frac12\bm\alpha_{\cal T}\bm h^{mn}) \nonumber\\
                                &\breve=& 0
\eeqn
%
where we used, in the first line, the fact that 
$\bm D^a\bm\alpha_{\cal T}$ is
a conformal Killing field and, in the second line, 
the facts that $\1 \bm E_{ab}$ is divergence-free and traceless.
In other words, the four-momentum is ``conserved'' in $\cal A$
because it consists of a
smooth conformal Killing field and a smooth divergence-free, 
traceless gravitational field.

Recall that there are {\em four} independent functions of
$\bm\alpha_{\cal T}$ as shown in eq.(5.6).
Thus, there are {\em four} independent components of {\em four}
momentum $\bm P$ as we expect.

Now let us evaluate the four-momentum defined above and see where the
information is contained. First, we evaluate the $\bm\alpha_t$
component. For simplicity, choose the $\chi=$const$.$ 
surface in $\cal A$ as $\cal S$. We obtain
\beqn
     \bm P(\bm\alpha_t) 
                   = \frac{1}{16\pi} \int_{\chi=\mbox{const$.$}}
                     \left( \frac{\partial^2}{\partial\bm\chi^2}\1 \bm F   
                            - \1 \bm F 
                     \right) \sinh\bm\chi \sqrt{\bm h}\ d\bm\theta d\bm\phi, 
\eeqn
where $\bm h := \mbox{det}(\bm h_{ij}) = \sinh^2\bm\chi \sin\bm\theta$.
Substituting the general solution of $\1\bm F$, eq.(\ref{eq:(1)F}),
and using the orthogonality of spherical harmonics
\beqn
     \int \bm Y_{\ell m} d(\cos\bm\theta)d\bm\phi = 
        4\sqrt 2\pi^{\frac{3}{2}} \ \delta_{\ell ,0} \ \delta_{m, 0},
\eeqn
we find that all the terms but $\1\bm F_{\ssr{00}}$ of 
$\1\bm F$ vanish and obtain
\beq
     \bm P(\bm\alpha_t) = \bm a_{00}.             \label{eq:P=a00}
\eeq
The spatial momentum can be evaluated likewise as 
\beqn
     & & \bm P(\bm\alpha_x) =  \bm a_{11}  \nonumber\\ 
     & & \bm P(\bm\alpha_y) =  \bm a_{1\mbox{-}1}   \\ 
     & & \bm P(\bm\alpha_z) =  \bm a_{10}. \nonumber 
\eeqn
Here, we find that the $\ell=0,1$ modes of $\1\bm F$
contain the information on the four-momentum of an {\sc afti-0} spacetime.
Since the function $\1\bm F$ is of the first order asymptotic
structure, the coefficients of $\1\bm F$ and thus the four momentum $\bm P$ is also 
of the first order asymptotic structure.
If we complete the Kerr spacetime and the Schwarzschild spacetime,
we obtain the expected answers for the four-momentum $\bm P$:
$\bm a_{00}$ equals the mass parameter $m$. See eqs.(\ref{eq:Fschw}) and
(\ref{eq:Fkerr}) for the explicit calculations.
Moreover, $\bm a_{11}$ equals $-mv$ in the leading order of $v$ 
for the Schwarzschild spacetime boosted by a constant velocity $v$ 
along the $x$-axis. See eq.(\ref{eq:FbooS}).

Because $\1\bm B_{ab}$ is also divergence-free and traceless 
and $\bm\xi\in{\cal L}_L$ is a conformal Killing field
of $(\bri,\bm h_{ab})$, we obtain a new ``conserved quantity'' 
if we replace $\1\bm E_{mn}\bm D^n\bm\alpha_{\cal T}$ 
in eq.(\ref{eq:defP}) with one of 
$\1\bm B_{mn}\bm D^n\bm\alpha_{\cal T}$, 
$\1\bm E_{mn}\bm\xi^n$ and $\1\bm B_{mn}\bm\xi^n$.
However, it is found that they all give exact 2-form integrands.
Hence, the ``conserved quantities'' identically vanish\cite{AR}.
Because there are no more gravitational fields in an {\sc afftis-1}
spacetime, four-momentum $\bm P$ can be considered as the sole ``conserved quantity'' of 
{\sc afftis-1} spacetimes.

Next, consider {\sc afftis-2} spacetimes. 
There exists a gravitational field $\2\bm B_{ab}$
that is divergence-free and traceless in such spacetimes.
Hence, we yield another ``conserved quantity'' by replacing 
$\1\bm E_{mn}\bm D^n\bm\alpha_{\cal T}$ 
in eq.(\ref{eq:defP}) with $\2\bm B_{mn}\bm\xi^n$.
This quantity can be regarded as angular momentum if a gauge condition 
$\1 B_{ab}\breve=0$ is imposed.
This is to reduce the Ti group to the Poincar\'e group so that 
the angular momentum is defined around an origin in the 
{\em four}-dimensional affine space
of the infinitesimal Ti translations\cite{AR}.
Note here that it is clarified that the gauge condition is needed for
the quantity to be defined around the four-dimensional origin, not for
the quantity to be ``conserved''. This point was left unclear in \cite{AR}

We conclude the section with a remark on how the conserved quantities
are associated with the hierarchy in the asymptotic structure. 
To define the conserved quantities, we needed vector fields that respect
the asymptotic symmetries and gravitational fields that are 
traceless and divergence-free.
However, the asymptotically flat spacetimes that satisfy 
only the weakest fall-off condition on the energy-momentum tensor
do not possess sufficient asymptotic structures 
that the spacetimes lack for the conserved quantities to be well defined.
As a result, four-momentum is defined for an {\sc afftis-1} spacetime
and angular momentum is defined for an {\sc afftis-2} spacetime.


\noindent

\section{summary and discussions}

In this paper, we have proposed a new definition of asymptotic flatness at timelike
infinity. Instead of rescaling the metric of a spacetime conformally
as in the previous works\cite{CC,JP}, we rescaled
3-metrics and normal vector fields to the time slices with different powers
to provide the information on the gravitational fields. 
As a result, the causal structure of the physical spacetime
is not manifest in the unphysical spacetime 
and the timelike infinity arises not as a point but as a {\em 3-manifold}.
The 3-manifold timelike infinity enables us 
to characterize the asymptotically flat spacetime under investigation
with a fall-off condition on the physical components of the
physical energy-momentum tensor. 
Hence, it is possible to discuss the properties of such a spacetime with
{\em the energy-momentum tensor falling off at a certain rate}
where the rate is specified with the function $\Omega$ 
that reads asymptotically the inverse of $t$, the observer's proper time.
This makes a sharp contrast with the treatment using the conformal
completion in which the spacetime is characterized with the awkward
differentiability conditions on the unphysical metric and thus such
an argument is not possible.
Hereafter in this section, we refer to the physical components of the physical
energy-momentum tensor simply by {\em the energy-momentum tensor} for brevity.
 
We found the following hierarchy in the asymptotic structure of
an asymptotically spacetime.
If the energy-momentum tensor falls off faster than the rate 
$\Omega^2$, the spacetime possesses an intrinsic structure called the zero-th order 
asymptotic structure. The structure consists of a function $F$ that
is unity on timelike infinity $\bri$, and the manifold $\bri$ with a
unit spacelike 3-hyperboloid metric $\bm h_{ab}$. 
The presence of the structure means that 
the spacetime is asymptotically Minkowskian.
An asymptotic symmetry group, called Ti group, can be defined to
preserve the structure. 
The group is found to be similar to the Poincar\'e group but differs in
that the 4-dimensional translation group is replaced with an infinite
dimensional additive group.
Further, by introducing the notion of
an asymptotic Killing field, it is found that such a spacetime
becomes asymptotically stationary at the rate $\Omega^2$ and 
in general does not become stationary at a faster rate. 
Although there exists an
asymptotically symmetry group, the spacetime is just asymptotically
Minkowskian and thus does not possess gravitational fields with
sufficient properties to have four-momentum defined.
If the energy-momentum tensor falls off faster than the rate $\Omega^3$,
there is enough higher order asymptotic structure for the four-momentum 
to be defined. The form of one of the asymptotic structure $\1 F$
is determined in such spacetimes. A conserved quantity associated with 
angular momentum can be defined only for spacetimes
with even more asymptotic structure: the energy-momentum tensor 
needs to fall off faster than the rate $\Omega^4$.
For the quantity to be regarded as angular-momentum
in the sense that the angular-momentum should be defined around the
{\em four} dimensional origin, 
the leading order of one of the gravitational fields, $\1 B_{ab}$, must vanish.


\bigskip
\centerline{\bf Acknowledgment} 
We would like to thank Katsuhiko Sato, Akio Hosoya and Yasushi Suto for their 
encouragements. We especially thank Akio Hosoya for his careful
reading of the manuscript.
We also thank Keita Ikumi for his fruitful discussion with us. 

\appendix
\section{some afftis spacetimes and some that are not}
We give the explicit construction of the completion of 
the Schwarzschild spacetime and show that it is an {\sc afftis} spacetime. 
Then, it is shown that the nature of the charge and the angular momentum of the
Kerr-Newmann spacetime appears in the second order asymptotic structure.
The function $F:=-\pounds_n \Omega$ which contains the information 
on four-momentum is explicitly calculated for a boosted Schwarzschild spacetime.
At the end, an example of spacetimes which do not satisfy the conditions of
{\sc afftis-$n$} spacetimes is given.

In static spherical coordinates,
the metric of Schwarzschild spacetime takes the form
\beq
     \hat{g}_{ab} = - ( 1 - 2m/r )( 1 - 2m/r ) (dt)_a (dt)_b
                    + ( 1 - 2m/r )^{-1}     (dr)_a (dr)_b 
                    + r^{2} (d\sigma)_{ab}.   \label{eq:sch}
\eeq
Introducing new coordinates $t = \Omega^{-1}\cosh\chi$ and $r = \Omega^{-1}\sinh\chi$, 
we obtain smooth $n^a$, $F$ and $q_{ab}$:
\beqn
n^a &=& - \frac{1+4m\Omega\sinh\chi-4m^2\Omega^2}
                      {1-2m\Omega/\sinh\chi}    (\partial_\Omega)^a
           - 4m \frac{\cosh\chi-m\Omega/\tanh\chi}
                       {1-2m\Omega/\sinh\chi}    (\partial_\chi)^a
                                                         \nonumber\\
         &=& -(\partial_\Omega)^a
             -[\Omega \ m \1 F_{00}+\Omega^2 \frac{4m^2}{\tanh^2\chi}+ O(\Omega^3)\ ]
                            (\partial_\Omega)^a     
            - [4m\cosh\chi + \Omega\frac{4m^2}{\tanh\chi}
            + O(\Omega^2)\ ](\partial_\chi)^a,                           \\
F &=& \frac{1+4m\Omega\sinh\chi-4m^2\Omega^2}
                      {1-2m\Omega/\sinh\chi}             
   = \hspace{1ex} 1 \hspace{1ex} + \Omega \ m \1 F_{00}
            + \Omega^2 \frac{4m^2}{\tanh^2\chi} + O(\Omega^3) \label{eq:Fschw}\\
& & \hspace{-10ex}\mbox{and} \nnb\\
     q_{ab} &=& 16 m^2 \frac{\cosh^2\chi -2m\Omega\cosh\chi/\tanh\chi
                             +m^2\Omega^2/\tanh^2\chi}
                            {1+2m\Omega(2\sinh\chi-1/\sinh\chi) 
                             - 12 m^2\Omega^2 +8 m^3\Omega^3/\sinh\chi}
                            (d\Omega)_a (d\Omega)_b           \nonumber\\
            & &\quad -4m\frac{\cosh\chi-m\Omega/\tanh\chi}
                  {1-2m\Omega/\sinh\chi} 
                [(d\Omega)_a(d\chi)_b + (d\chi)_a(d\Omega)_b] \nonumber\\
            & &\quad +\frac{1+4m\Omega\sinh\chi-4m^2\Omega^2}
                      {1-2m\Omega/\sinh\chi} (d\chi)_a(d\chi)_b
             +  \sinh^2\chi(d\sigma)_{ab}               \nonumber\\
            &=& h_{ab} 
                 + [ 16m^2\cosh^2\chi
                 -\Omega \ 64\frac{m^3\cosh\chi\sinh^2\chi}{\tanh\chi}
                 +O(\Omega^2)](d\Omega)_a (d\Omega)_b         \nnb\\
            & &\hspace{3.7ex}
                 -[4m\cosh\chi + \Omega\frac{4m^2}{\tanh\chi} 
                      + O(\Omega^2)\ ](d\chi)_{{\ssr(}a}(d\Omega)_{b\ssr)}
                 +[\Omega \ 2m \1 F_{00} + \Omega^2\frac{4m^2}{\tanh\chi} 
                    + O(\Omega^3)\ ] (d\chi)_a(d\chi)_b 
\eeqn
Together with the fact that it is a vacuum solution of the Einstein equation,
the above calculations show that the Schwarzschild spacetime is an {\sc afti-}$\infty$
spacetime, where we denote a spacetime that is an {\sc afti-}$n$ spacetime for arbitrary
$n$ by an {\sc ati-}$\infty$ spacetime.
 Now take the outside of the event horizon $r>2m$ 
in the original coordinates as $\hat{\cal F}$.
Because the region $r\leq 2m$ corresponds to the point $\chi=0$ on the
$\Omega=0$ surface,
the choice provides timelike infinity $\bri$ with the point missing. 
Thus, topology of $\bri$ is $S^2\times(\real-\{p\})$ and
the Schwarzschild spacetime satisfies the condition of 
an {\sc afftis-}$n$ spacetime to arbitrary $n$.
It is important to note here that the Oppenheimer-Snyder spacetime, 
which describes dust collapsing to form a Schwarzschild black hole,
can be completed to possess $\bri$ with the same structure 
as the above completed Schwarzschild spacetime does.
The spacetime has the Schwarzschild metric outside the dust and
the surface of the dust collapses along 
$r=(r_{ini}/2)(1+\cos\eta)$ where $r$ and $r_{ini}$ are 
the Schwarzschild radial coordinate and the initial dust radius
at time $t=0$, respectively; 
and $\eta$ is related to the Schwarzschild time coordinate $t$
by
\beq
      t=2m\ln\left| \frac{\sqrt{r_{ini}/2m-1}+\tan{\eta/2}}
                       {\sqrt{r_{ini}/2m-1}-\tan{\eta/2}} \right|
       +2m\sqrt{r_{ini}/2m-1}(\eta+(\eta+\sin\eta)r_{ini}/4m).
\eeq
If we choose the region $r>r_{ini}$ as region $\h{\cal F}$,
the metric of the region is that of the Schwarzschild spacetime and thus
$\h{\cal F}$ satisfies all the conditions of {\sc afftis-}$n$ for
arbitrary $n$ and can be completed to have exactly 
the same $n^a$, $F$ and $q_{ab}$ in $\cal F$ as those of the Schwarzschild 
spacetime. 
Since the region $r\leq r_{ini}$ corresponds to the point $\chi=0$ on
$\bri$ in this completion also, 
the global structure of $\bri$ is also the same as that of the 
above completed Schwarzschild spacetime.

In static spherical coordinates, the metric of Kerr-Newmann spacetime is given by
\beqn
     \h g_{ab} &=& 
       - \frac{\Delta-a^2\sin^2\theta}{\Sigma}(dt)_a(dt)_b
       -2\frac{a\sin^2\theta(r^2+a^2-\Delta)}{\Sigma}(dt)_{{\ssr(}a}(d\phi)_{b\ssr)}
          \nnb\\& &
       +\frac\Sigma\Delta(dr)_a(dr)_b+ \Sigma(d\theta)_a(d\theta)_b
       +\frac{(r^2+a^2)^2-\Delta a^2\sin^2\theta}{\Sigma}\sin^2\theta(d\phi)_a(d\phi)_b
\eeqn
where $\Sigma=r^2+a^2\cos^2\theta$ and $\Delta=r^2+a^2+e^2-2mr$.
Introducing $\eta:=\ln\Omega$, we clearly see that
the nature of the charge and the angular momentum of the spacetime
appear in the second order asymptotic structure and do not alter the 
lower order asymptotic structure:
\beqn
    n^a &=& e^{-\eta}[\upp0\,n^a + \Omega  \,\upp1\,n^a 
                     + \Omega^2\,\upp2\,n^a +O(\Omega^3)]       \nnb\\
    F\, &=& \,\,\up{(0)}F\,\, + \Omega \,\,\1 F\,\, 
            + \Omega^2 \,\,\2 F\,\, +O(\Omega^3)\nnb\\
 q_{ab} &=& \upp0\,q_{ab} + \Omega \,\upp1\,q_{ab} 
                          + \Omega^2\,\upp2\, q_{ab} +O(\Omega^3)
\eeqn
where
\beqn
  \upp0\, n^a &=& -(\ppp_\eta)^a, \hspace{8ex}
  \upp1\, n^a  =  m \1 F_{00} (\ppp_\eta)^a - 4m\cosh\chi(\ppp_\chi)^a 
                                \hspace{8ex}\mbox{and} \nnb\\
  \upp2\, n^a &=& (\frac{4m^2-e^2}{\tanh\chi}-e^2-a^2\sin^2\theta)
               (\ppp_\eta)^a
               -\frac{4m^2-2e^2-a^2\sin^2\theta}{\tanh\chi}(\ppp_\chi)^a
                +2ma\frac{\cosh\chi}{\sinh^3\chi}(\ppp_\phi)^a;
\eeqn
\beqn
    \up{(0)}F &=& 1, \hspace{8ex} 
         \1 F = m \1 F_{00}  \hspace{8ex} \mbox{and} \nnb\\
         \2 F &=& \frac{4m^2-e^2}{\tanh\chi}-e^2-a^2\sin^2\theta;
    \label{eq:Fkerr}
\eeqn
\beqn
    \upp0\,q_{ab} &=& h_{ab},
      \hspace{8ex}
    \upp1\,q_{ab}  =  -2(4m\cosh\chi)(d\eta)_{{\ssr(}a}(d\chi)_{b\ssr)} 
                      + 2m \1 F_{00}(d\chi)_a(d\chi)_b 
      \hspace{8ex} \mbox{and} \nnb\\
    \upp2\,q_{ab} &=&  16m^2 \cosh^2\chi(d\eta)_a(d\eta)_b
                     - 2\frac{4m^2-2e^2-a^2\sin\theta}{\tanh\chi}
                         (d\eta)_{{\ssr(}a}(d\chi)_{b\ssr)} 
                     + 4ma\sin^2\theta
                  (\frac{(d\eta)_{{\ssr(}a}(d\phi)_{b\ssr)}}{\tanh\chi}
                        - (d\chi)_{{\ssr(}a}(d\phi)_{b\ssr)} ) \nnb\\
                  & &+(\frac{4m^2-e^2-a^2\sin\theta}{\tanh^2\chi}-e^2)
                         (d\chi)_a(d\chi)_b     
                     + a^2\cos^2\theta (d\theta)_a(d\theta)_b
                     + a^2\sin^2\theta (d\phi)_a(d\phi)_b.
\eeqn

In the boosted Schwarzschild spacetime with velocity of $v$ along
$x-$axis, or to be precise in the Schwarzschild spacetime that is 
observed by an observer with velocity of $v$ along $x$-axis
\beqn
     F &=& 1 + \Omega \lnk 2m (2\sinh\chi + \frac{1}{\sinh\chi}) -2mv
              (2\cosh\chi - \frac{\cosh\chi}{\sinh^2\chi})\sin\theta \cos\phi  
                    \rnk  + O (\Omega^2) \nonumber\\
       &=& 1 + \Omega \lnk \ m\1 F_{00}  -mv\1 F_{11} + O(v^2)
                    \rnk  + O (\Omega^2). \label{eq:FbooS}
\eeqn

Let us consider the flat Friedmann-Robertson-Walker spacetime,
which does not qualify as an {\sc afti} spacetime
as one expects.
The metric of the spacetime filled with perfect fluid is
\beqn
     \hat{g}_{ab} = - (dt)_a (dt)_b + a^2(t)[ (dr)_a(dr)_b  
                               + r^2(d\sigma)_{ab} ],     \nonumber
\eeqn
where $a(t) = a_0(t/t_0)^{2/3(1+w) }$
if the energy density $\rho$ and the pressure $p$ is related by $p=w\rho$. 
In order to get smooth $n^a$ and $q_{ab}$,
we need to choose the new coordinates $(\Omega,\chi)$ as 
\beqn
   t &=& a_0\frac{1+3w}{1+w}\,\Omega^{-1} \,\,
                (\cosh\chi)^{\frac{3+3w}{1+3w}} \nonumber\\ 
   r &=& \,\,t_0^{\frac{2}{3+3w}}\,\,\Omega^{-\frac{1+3w}{3+3w}}\,\,
                        (\sinh\chi).\nonumber
\eeqn
They give us
\beqn
     n^a &=& \frac1{a_0^2}\left( \frac{3+3w}{1+3w} \right)^2
               (\cosh\chi)^{-\frac{4}{1+3w}} ( \partial_\Omega )^a \\
     q_{ab} &=& \,a_0^2 \,(\cosh\chi)^{\frac{4}{1+3w}}[ (d\chi)_a(d\chi)_b  
              + \sinh^2\chi (d\sigma)_{ab} ].               
\eeqn
However, with this choice of $\Omega$,
the time-time component of the energy-momentum tensor decays as $\Omega^2$:
\beqn
      \h n^a\h n^b\hat{T}_{ab} 
                &=& 9\rho_0 \left(  \frac{t}{t_0}  \right)^{-2}
                        \cosh^2\chi + O(\Omega)\nonumber\\
                &=& \Omega^2 \rho_0 
                    \left(   \frac{3t_0(1+w)}{a_0(1+3w)}  \right)^2 
                    (\cosh\chi)^{-\frac4{1+3w}} + O(\Omega^3)\nonumber
\eeqn
from which we see that the spacetime does not satisfy the third condition of 
{\sc afti-$n$} spacetimes for non-negative $n$ and 
thus the spacetime is not an {\sc afti-$n$} spacetime.
Hence, the spacetime does not need to possess the properties of an 
{\sc afti-$n$} spacetimes: the function 
\beqn
     F \breve=\frac1{a_0^2}\left( \frac{3+3w}{1+3w} \right)^2
               (\cosh\chi)^{-\frac{4}{1+3w}},
\eeqn
is not unity on $\bri$; and
the spacetime does not admit asymptotic stationary Killing field.


\end{document}